\documentclass[runningheads]{llncs}
\usepackage[T1]{fontenc}
\usepackage{amsmath}
\usepackage{amsfonts}
\usepackage{hhline}
\usepackage{hyperref}
\usepackage{multirow}
\usepackage{graphicx}
\usepackage{epstopdf}
\usepackage{amssymb}
\usepackage{array}
\usepackage{xcolor}
\usepackage{cite}
\DeclareMathAlphabet{\mathcal}{OMS}{cmsy}{m}{n}
\usepackage[mathscr]{eucal}
\DeclareMathAlphabet\mathbfcal{OMS}{cmsy}{b}{n}
\usepackage{multirow}
\usepackage[ruled,vlined,linesnumbered]{algorithm2e}
\usepackage{amsmath}

\usepackage[font={small}]{caption}
\usepackage[font={footnotesize}]{subcaption}
\usepackage{cancel}
\usepackage[flushmargin]{footmisc}

\let\oldnl\nl
\newcommand{\nonl}{\renewcommand{\nl}{\let\nl\oldnl}}

\usepackage{fancyhdr}
\fancypagestyle{firststyle}
{
   \fancyhf{}
   \fancyfoot[C]{\footnotesize \thepage}
   \fancyhead[C]{\small European Symposium on Research in Computer Security (ESORICS) 2022}
}

\usepackage{bbm} 
\usepackage{enumitem}

\newcommand{\sref}[1]{Section~\ref{#1}}

\begin{document}

\title{Long-Short History of Gradients is All You Need: Detecting Malicious and Unreliable Clients in Federated Learning}
\author{Ashish Gupta\inst{1} \and Tie Luo\inst{1}
\and Mao V. Ngo\inst{2} \and Sajal K. Das\inst{1}}

\authorrunning{Gupta et al.}
\titlerunning{Long-Short History of Gradients is All You Need}
\institute{Missouri University of Science and Technology, Rolla, USA \and Singapore University of Technology and Design, Singapore \\
\email{ashish.gupta@mst.edu, tluo@mst.edu,\\ vanmao\_ngo@sutd.edu.sg, sdas@mst.edu}}
 \maketitle

\thispagestyle{firststyle}

\begin{abstract}\vspace{-7mm}
Federated learning offers a framework of training a machine learning model in a distributed fashion while preserving privacy of the participants. As the server cannot govern the clients’ actions, nefarious clients may attack the global model by sending malicious local gradients. In the meantime, there could also be {\em unreliable} clients who are {\em benign} but each has a portion of low-quality training data (e.g., blur or low-resolution images), thus may appearing similar as malicious clients. Therefore, a defense mechanism will need to perform a {\em three-fold} differentiation which is much more challenging than the conventional (two-fold) case. This paper introduces MUD-HoG, a novel defense algorithm that addresses this challenge in federated learning using {\em long-short history of gradients}, and treats the detected malicious and unreliable clients differently. Not only this, but we can also distinguish between {\em targeted} and {\em untargeted attacks} among malicious clients, unlike most prior works which only consider one type of the attacks. Specifically, we take into account sign-flipping, additive-noise, label-flipping, and multi-label-flipping attacks, under a non-IID setting. We evaluate MUD-HoG with six state-of-the-art methods on two datasets. The results show that MUD-HoG outperforms all of them in terms of accuracy as well as precision and recall, in the presence of a mixture of multiple (four) types of attackers as well as unreliable clients. Moreover, unlike most prior works which can only tolerate a low population of harmful users, MUD-HoG can work with and successfully detect a wide range of malicious and unreliable clients - up to $47.5\%$ and $10\%$, respectively, of the total population. Our code is open-sourced at {\tt\small https://github.com/LabSAINT/MUD-HoG\_Federated\_Learning}.
\end{abstract}

\section{Introduction}
\vspace{-0.1in}
In recent years, the proliferation of smart devices with increased computational capabilities have laid a solid foundation for training machine learning (ML) models over a large number of distributed devices. Traditional ML approaches require the training data to reside at a central location; the distributed ML case requires a well-controlled data-center-like environment. Such approaches demand high network bandwidth and provoke great privacy concerns. To this end, Google introduced the concept of Federated Learning (FL)~\cite{mcmahan2017communication} which allows distributed clients to collaboratively train a global ML model without letting their data leave the respective devices. At a high level, it works as follows. A central server initiates the training process by disseminating an initial global model to a set of clients. Each client updates the received model using its local data and sends back the updated model (not data). The server aggregates the received model updates (weights or gradients) into a global model and disseminates it again back to the clients. This procedure repeats until the global model converges. FL is advantageous in preserving data privacy and saving communication bandwidth, and has been applied to a wide range of applications in the Internet of Things (IoT)~\cite{khan2021federated}, natural language processing~\cite{hard2018federated,leroy2019federated}, image processing~\cite{liu2020fedvision}, etc. 

However, the uncontrolled and distributed nature of the clients, as well as the server's inaccessibility to clients' data, make FL vulnerable to adversarial attacks launched by clients~\cite{wu2020federated, bhagoji2019analyzing, bagdasaryan2020backdoor, mao2021romoa,awan2021contra}. In general, a malicious client (adversary) can launch two types of attacks: (1) an \textit{untargeted attack}, sometimes referred to as a Byzantine attack~\cite{cao2021fltrust,li2019rsa, wu2020federated}, where the adversary attempts to corrupt the {\em overall} performance of the global model (e.g., degrade a classifier's accuracy on all classes); (2) a \textit{targeted attack}, where the adversary aims to degrade the model performance only for some specific cases (e.g., misclassify all dogs to cat) while not affecting the other cases~\cite{fung2020limitations,mao2021romoa}. 
Untargeted attacks could be tackled by robust aggregation techniques~\cite{blanchard2017machine, xie2018generalized, chen2017distributed} when data are independent and identically distributed (IID) among the clients, whereas targeted attacks are much harder to defend because their specific targets are often unknown to the defender. 

Another category of clients, which are largely overlooked in the FL security literature, are {\em unreliable clients}. These are benign clients but some of their data are of low quality and hence may appear as if their model updates were malicious too. For example, IoT devices such as sensors, smartphones, wearables, and surveillance cameras, are often subject to rigid hardware limitations and harsh ambient environments and thus may produce low-quality and noisy data~\cite{jiang2020federated}. A simplified solution could be one that treats clients who do not improve classification performance over a number of rounds as unreliable, and excludes them from aggregation in subsequent rounds, like in \cite{ma2021federated,mallah2021untargeted}. However, firstly this does not differentiate between benign and malign clients; secondly, excluding unreliable clients is not always desirable because such clients may possess valuable data such as infrequent classes on which other clients have no or few samples.

In this paper, we tackle the challenge of detecting and distinguishing between malicious and unreliable clients, as well as between targeted and untargeted attackers (among malicious clients), in FL. The main idea of our approach is to use {\em long-short history of gradients} jointly with judiciously chosen distance and similarity metrics during the iterative model updating process. Unlike prior works in \cite{fung2020limitations,mao2021romoa,awan2021contra,blanchard2017machine,chen2017distributed} which only consider attackers, we identify unreliable clients and take advantage of their contributions. We further consider both targeted and untargeted attacks and more fine-grained attack types: (untargeted) additive-noise and sign-flipping attacks, and (targeted) single- and multi-label-flipping attacks. Moreover, unlike prior works in \cite{blanchard2017machine, xie2018generalized, chen2017distributed}, we consider non-IID data settings which are more representative of real-world FL scenarios with heterogeneous clients.

The main contributions of this paper are summarized as follows:
\begin{itemize}
\vspace{-0.1in}
\item We propose a novel approach MUD-HoG that stands for {\textbf{M}alicious and \textbf{U}nreliable Client \textbf{D}etection using \textbf{H}istory \textbf{o}f \textbf{G}radients}. To the best of our knowledge, this is the first work that detects both malicious attackers and unreliable clients in FL, distinguishing between targeted and untargeted attackers. It allows the server to treat the clients in a more {\em fine-grained} manner, by exploiting unreliable clients' low-quality (but still useful) data.
\item We introduce {\em short HoG} and {\em long HoG} and a sequential strategy that uses them in a carefully-designed way, allowing us to achieve the above goal. In addition, we achieve our goal in a non-IID setting which is more realistic and challenging, with the presence of mixed types of attackers.

\item We conduct extensive experiments to evaluate MUD-HoG in terms of accuracy, precision, recall, and detection ratio, on two benchmark datasets in comparison with 6 prior FL security mechanisms. The results show that MUD-HoG withstands up to 47.5\% clients being malicious with a negligible ($\sim$1\%) compromise of accuracy, and comprehensively outperforms all the baselines on the considered metrics.
\end{itemize}
\vspace{-0.05in}
The rest of the paper is organized as follows. Section~\ref{related-work} reviews the related literature while Section~\ref{problem_setting} define the problem statement with the types of clients and considered attacks. Section~\ref{solution} presents the proposed MUD-HoG approach with novel concepts of short HoG and long HoG, and Section~\ref{sec:ExpEvaluation} evaluates the robustness of the approach by conducting extensive experiments. Finally, Section~\ref{conclusion} concludes the paper with future research directions.

\vspace{-0.15in}
\section{Related Work}\label{related-work}
\vspace{-0.1in}
\subsection{Distributed ML with Malicious Clients}
\vspace{-0.1in}
Defending against malicious clients has been explored in distributed ML~\cite{yin2018byzantine, blanchard2017machine, xie2019zeno}. It has been noted that the stochastic gradient descent (SGD) algorithm is vulnerable to untargeted (Byzantine) attacks where malicious clients send random/arbitrary gradients to the server to negatively affect the convergence or performance of the global model. Methods such as Krum and Multi-Krum~\cite{blanchard2017machine}, Medoid~\cite{xie2018generalized}, and GeoMed~\cite{chen2017distributed} have been proposed to defend against Byzantine attacks by extending SGD with a robust aggregation function. In another work~\cite{sun2019can}, the authors argued that the effect of malicious clients can be mitigated by gradient or norm clipping based on a threshold assuming that the attacks produce boosted gradients.
However, these methods assume IID data, which often does not hold in FL settings. In addition, they aim to {\em tolerate} malicious clients rather than {\em distinguishing} them from normal ones, and thus may lead to cumulative negative impact over time and is also less preferable.
  
\vspace{-0.15in}
\subsection{FL under Untargeted Attacks}
\vspace{-0.1in}
Various Byzantine-robust algorithms have been proposed for FL's non-IID settings in recent years. For example, a class of subgradient-based algorithms is proposed to defend malicious clients by robustifying the objective function with a regularization term~\cite{li2019rsa}. However, these algorithms only consider simple attacks such as same-value and sign-flipping attacks. In another work~\cite{wu2020federated}, a variance reduction scheme inherited from~\cite{defazio2014saga} is combined with model aggregation to tackle untargeted attacks. In~\cite{cao2021provably}, the authors provided provable guarantees to ensure that the predicted label of a testing sample is not affected by the attack. They also proposed an ensemble method with a voting strategy to address the case of a bounded number of malicious clients. 
However, similar to some of the works discussed in the distributed ML case, this ensemble method cannot identify which clients are malicious. The above Byzantine-robust algorithms fail to stand against the attackers if they are present in high percentage. Moreover, all the above works are vulnerable to targeted attacks such as label flipping~\cite{tolpegin2020data}.

\vspace{-0.15in}
\subsection{FL under Targeted Attacks}
\vspace{-0.1in}
As targeted attacks aim to reduce the model performance only on certain tasks while maintaining a good performance on others, they are elusive and harder to detect~\cite{mao2021romoa}. One of the popular defense methods, called FoolsGold~\cite{fung2020limitations}, attempts to detect targeted attackers (e.g., label-flipping) based on the diversity of client contributions over the training rounds with an unknown number of attackers. With more realistic FL settings, Awan {\em et al.}~\cite{awan2021contra} also exploited the clients' per-round contribution and cosine-similarity measure to defend against data poisoning attackers. In~\cite{li2020learning}, an anomaly detection framework is proposed to differentiate anomalous gradients from normal ones in a low-dimensional embedding (spectral) using reconstruction errors. 
However, it requires a pre-trained model on a reference dataset at the server prior to start the training process, which is a strong requirement often not met in FL settings.
Mao \textit{et al.}~\cite{mao2021romoa} treated FL as a repeated game and introduced a robust aggregation model to defend against targeted and untargeted adversaries by designing a lookahead strategy based similarity measure. However, like many studies discussed earlier, it tolerates but does {\em not distinguish} adversaries from normal clients.
Moreover, since most existing works~\cite{li2020learning,mao2021romoa,fung2020limitations,awan2021contra} consider only two types of clients (normal and malicious), they may treat an unreliable client (who possesses lower-quality data) as malicious, which is not desirable. 

In this work, we do not include {\em backdoor attacks} \cite{ozdayi2021defending, xie2021crfl,wang2020attack,awan2021contra}, 
which are a sub-category of data poisoning attack triggered by a particular pattern (e.g., pixel patch) embedded into data (e.g., images). However, unlike prior work, we include unreliable clients which are more likely to encounter in realistic FL deployments.

We also highlight that the term {\em unreliable} or {\em irrelevant} clients used in some studies~\cite{ma2021federated,mallah2021untargeted, nagalapatti2021game} means clients whose contributions do not make any progress (i.e., improve model accuracy) over the past few rounds, which is considerably different from our definition of unreliable clients (see \sref{sec:client_types}) which refers to clients who have low-quality data.

\vspace{-0.15in}
\section{Model}\label{problem_setting}
\vspace{-0.1in}
We consider a typical FL framework with a central server and multiple clients participating in a collaborative model training process for a classification task using a deep learning model.

\vspace{-0.15in}
\subsection{FL Preliminaries}\label{sec:fl}
\vspace{-0.1in}
Let $N$ be the total number of clients participating in the FL model training process. Out of these $N$ clients, $m$ of them are malicious, and $u$ of them are unreliable. Thus, there are $n = N-m-u$ normal clients. 
We consider a typical FL scenario for building a neural network model, where all clients share a common model structure under the same learning objective. The server initiates training by sending a global model $\boldsymbol{w}$ (e.g., random weights) to all clients. Each client updates the model $\boldsymbol{w}$ by training on its local dataset a certain number of epochs, and sends back the updated gradients. Note that sending gradients is equivalent to sending model parameters (weights). During training, each client learns the new weights $\boldsymbol{w}'$ by minimizing a loss function $\mathcal{L}(h_w(x),y)$ (e.g., cross-entropy loss function) over multiple epochs, where the function $h_w(\cdot)$ maps input data samples $x$ to labels $y$. At a round $\tau$, a client $c_i$ computes the gradients as follows:
\begin{align}\label{grad}
 \nabla_{\tau, i} = \boldsymbol{w}_\tau - \underset{\boldsymbol{w}}{\text{argmin}} \quad \mathcal{L}(h_{i,\boldsymbol{w}}(x),y).
\end{align}

Let the client $c_i$ hold a local dataset $\mathcal{D}_i$ which can be non-IID as compared to other clients. When all clients are normal, the server aggregates all the gradients received from the clients, by
\begin{align} \label{agg}
 \boldsymbol{\nabla}_\tau = \sum_{i=1}^{N} \frac{|\mathcal{D}_i|}{|\boldsymbol{\mathcal{D}}|}  \nabla_{\tau,i},
\end{align}
where $|\boldsymbol{\mathcal{D}}| = \sum_{i=1}^{N} |\mathcal{D}_i|$. The weights of the global model for the next round $\tau+1$ are then updated as $\boldsymbol{w}_{\tau+1} = \boldsymbol{w}_{\tau} - \eta \boldsymbol{\nabla}_\tau$, where $\eta$ is the learning rate. 

\vspace{-0.15in}
\subsection{Client Types}\label{sec:client_types}
\vspace{-0.01in}
For generality, we consider a heterogeneous FL setting in which clients may be sensor boards, smartphones, surveillance cameras, laptops, connected vehicles, etc., owned by individuals or organizations. As a result, their data could be non-IID and thus each client could contribute to the global model training. 
We consider three types of clients and the last is further categorized in terms of attack types (see \sref{sec:threat}) the malicious client can launch.
\begin{enumerate}[label=\arabic*)]
 \item \textit{Normal clients} honestly participate in the model training process and have good-quality data.
 
 \vspace{3pt}
 \item \textit{Unreliable clients} participate honestly in the FL but have some of its data are of low-quality. 
 These data, however, could be exploited to improve diversity, especially if they capture distributions that normal clients fail to (or inadequately do). For example, A low-end camera does not produce high-resolution images but may capture some infrequent classes of images that other clients do not. Note that our definition of ``unreliable client'' is different from that in ~\cite{ma2021federated,mallah2021untargeted} and also from the ``irrelevant client'' in~\cite{nagalapatti2021game}, where they mean a client who does not make progress (i.e., improve model accuracy) over the past few FL rounds, which therefore is a useless client.

\vspace{3pt}
 \item \textit{Malicious clients} are attackers who manipulate their local training data (i.e., {\em data poisoning}) or model weights/gradients (i.e., {\em model poisoning}) to generate adversarial impact on the global model being trained. For example, they may alter the labels of some of their data samples or perturb their local gradients before sending to the server.  
 
\end{enumerate}

\begin{figure}[h]
\vspace{-0.25in}
\centering
\includegraphics[scale=0.9]{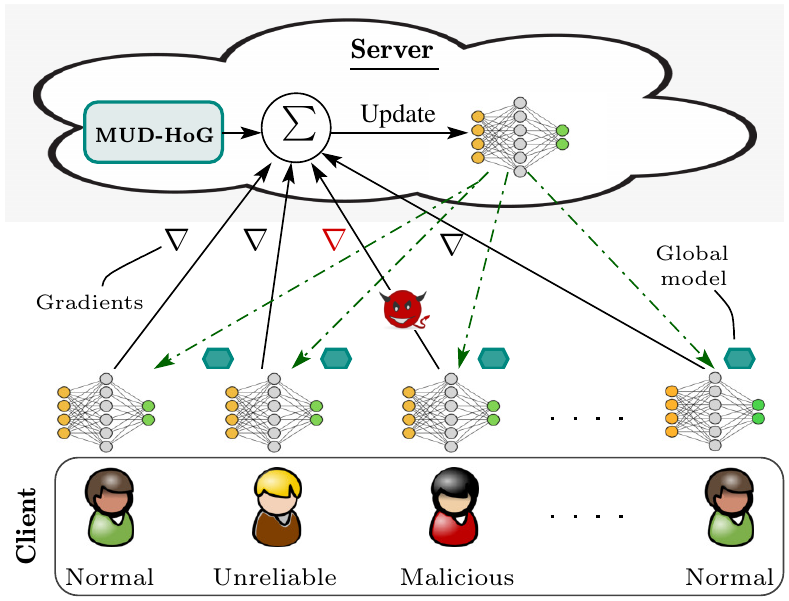}
\caption{Overview of FL with mixed types of clients. Malicious clients include targeted and untargeted attackers.} \label{clients}
\vspace{-0.2in}
\end{figure}

With the presence of mixed types of clients having non-IID data, our problem is more realistic and challenging than prior work such as \cite{fung2020limitations,awan2021contra,chen2017distributed}. Fig.~\ref{clients} provides an overview of our problem setting, where MUD-HoG runs at the server.

\noindent \textbf{Problem statement.} The problem in hand is two-fold: (1) How to identify and differentiate malicious clients (together with their attacks) from unreliable clients at the server while performing model aggregation? (2) How to mitigate the negative influence of malicious clients on the global model while still taking advantage of unreliable clients' updates? Let us reformulate Eq.~(\ref{agg}) as:

\begin{align}\label{eq:agg_unrl}
 \boldsymbol{\nabla}_\tau = \sum_{i\in \mathcal{C}_{norm}} \frac{|\mathcal{D}_i|}{|\boldsymbol{\mathcal{D}}|} \nabla_{\tau,i} 
+ \alpha \sum_{i\in \mathcal{C}_{unrl}} \frac{|\mathcal{D}_i|}{|\boldsymbol{\mathcal{D}}|} \nabla_{\tau,i},
\end{align}
where $\mathcal{C}_{norm}$ and $\mathcal{C}_{unrl}$ are the set of normal clients and that of unreliable clients, respectively, and the parameter $\alpha\in(0,1)$ down-weights the gradients of unreliable clients. Note that malicious clients are excluded.

\vspace{-0.1in}
\subsection{Threat Model}\label{sec:threat}
\vspace{-0.1in}
A malicious client can launch either of the following attacks:\vspace{-2mm} 
\begin{itemize}
\item {\bf Untargeted attack.} The objective here is to downgrade the {\em overall} performance of the global model. The following two model poisoning attacks are considered: (i) {\em Sign-flipping.} The malicious client flips the sign of its local gradients (from positive to negative and vice versa) before sending them to server, while the magnitude of the gradients remains unchanged.  
(ii) {\em Additive-noise.} The malicious client adds Gaussian or random noise to its local gradients before sending to the server.  

\item {\bf Targeted attack.} The objective is to decrease model performance on particular cases while not affecting other cases. The following two data poisoning attacks are considered: (i) {\em Label-flipping.} The attacker changes the label of all the instances of one particular class (source label), say $y_1$, to another class (target label), say $y_2$, while (intentionally) keeping other classes intact to avoid being detected. (ii) {\em Multi-label-flipping.} The attacker flips multiple source labels to a particular target label. This will result in the target label has an increased accuracy while harming the accuracy on other classes.
\end{itemize}

We make the following {\em Assumptions}: (i) Each attacker can only manipulate its own data or model but not other clients' or modify the server's aggregation algorithm. (ii) Number of malicious clients (including untargeted and targeted attackers) is less than other clients (including normal and unreliable). (iii) Malicious clients are persistent, meaning that they attack in every round.

\vspace{-0.15in}
\section{MUD-HoG Design}
\label{solution}
\vspace{-0.1in}

MUD-HoG runs at the server to defend the global model. Unlike existing work such as \cite{blanchard2017machine}, MUD-HoG assumes that the number of malicious clients is {\em unknown} to the server.

\vspace{3pt}
\noindent \textbf{Challenges.} The design challenges come from the following factors: the mixed types and unknown distribution of clients, non-IID data, and the server's inaccessibility to client data. The only information that the server has is the gradients (Eq.~\ref{grad}) sent by the clients each round, as a result of their local optimization such as stochastic gradient descent (SGD) over the loss function $\mathcal{L}(\cdot)$.

With targeted attacks, the malicious clients share a common objective and thus will have similar gradients~\cite{fung2020limitations} between each other. On the other hand, gradients from untargeted attackers would be dissimilar from each other since they perturb gradients randomly or flip gradient signs. This gradient space is rather complex and irregular, insofar as there is no single appropriate similarity measure that can distinguish malicious clients from the normal ones. Furthermore, unreliable clients introduce another degree of complication as they would behave very similar to {\em untargeted} attackers and hence are hard to distinguish.

\vspace{3pt}
\noindent\textbf{Long-Short History of Gradients (HoG).} 
We propose two new notions of HoG, based on which we design a robust algorithm MUD-HoG 
to address the above challenges. Let $\nabla_i = \{\nabla_{1,i}, \nabla_{2,i}, \cdots, \nabla_{\tau-1,i} \}$ denote the collection of HoGs received by the server from client $c_i$ prior to the $\tau^{th}$ round.

\vspace{3pt}
\noindent \textbf{Definition 1 (Short HoG).} The short HoG of client $c_i$ at round $\tau$, defined as,
\begin{align}
\nabla^{sHoG}_i = \frac{1}{l}\sum_{t=\tau-l}^{\tau-1} \nabla_{t,i}    
\end{align}
is a moving average of $c_i$'s gradients of the last $l$ rounds, where $l$ is the sliding window size. The short HoG smooths a client's gradients to remove single-round randomness.

\vspace{3pt}
\noindent \textbf{Definition 2 (Long HoG).} The long HoG of client $c_i$ at round $\tau$ is defined as
\begin{align}
\nabla^{lHoG}_i = \sum_{t=1}^{\tau-1} \nabla_{t,i},
\end{align}
which is the sum of all the gradients in the set $\nabla_{i}$. Thus, the long HoG captures the {\em accumulated} influence of a client on the global model, which reflects its goal. 

Note that, at any round $\tau$, the server does not need to store all the previous gradient vectors $\{\nabla_{1,i}, \nabla_{2,i}, \cdots, \nabla_{\tau-1,i} \}$ received from the client $c_i$; instead, it only needs to keep $l$ latest vectors for computing short HoG and the {\em sum} of all the previous vectors for long HoG. Hence, at each round, the server would keep only $l+1$ gradient vectors for each client. Therefore, the required memory is independent of the number of training rounds $\tau$, and one should not have memory concerns when $\tau$ increases.

\vspace{-0.1in}
\subsection{Sequential Strategy}
\begin{figure}[h]
\vspace{-0.15in}
 \centering
 \includegraphics[scale=1.1]{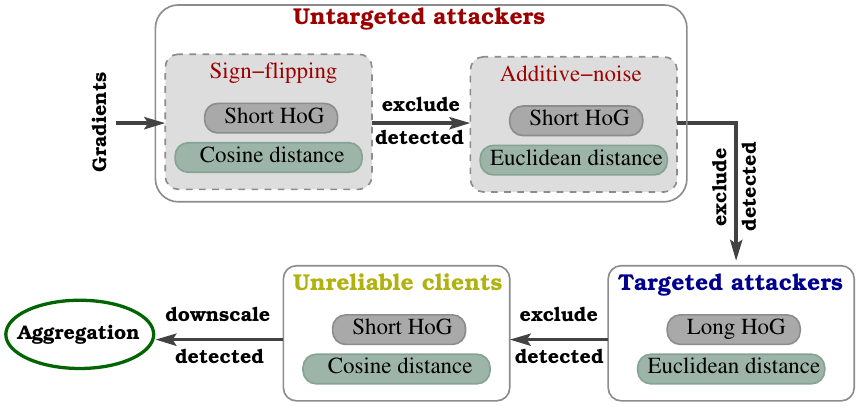} 
 \caption{Overview of MUD-HoG with the gradient space (short or long HoG) and similarity measures (Euclidean or cosine) used for detecting different types of clients.}
 \label{overview}
\vspace{-0.2in}
 \end{figure}

\vspace{-0.1in}
By introducing short HoG and long HoG, MUD-HoG exploits two different gradient space and follows a sequential strategy to detect the type of each client in the following order: untargeted, targeted, unreliable, and normal, as depicted in Fig.~\ref{overview}. The key ideas are discussed in the following steps.

\begin{figure}[h]
\vspace{-0.25in}
\centering
\includegraphics[scale=1.5]{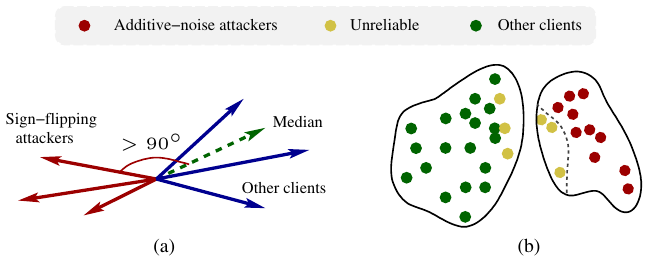} 
\vspace{-0.1in}
\caption{Illustration of (a) the angular deviation of sign-flipping attackers from the median client (green), and (b) clustering of additive-noise attackers, unreliable clients, and other clients after excluding sign-flipping attackers.}
\label{signflip}
\vspace{-0.2in}
\end{figure}

\noindent \textbf{1) Untargeted attack.} We can deduce the untargeted intention from the client's short HoG. Since an untargeted attacker aims to corrupt the whole model, for example using sign flipping or additive noise, its short HoG would differ substantially from normal clients. 
First, in the case of sign-flipping attack, a malicious client essentially changes its gradient to the opposite direction, which would result in a large angular deviation 
from the {\em median} gradient of all the clients, as depicted in Fig.~\ref{signflip}a. This also justifies that using {\em cosine distance} in the space of short HoG would be an appropriate choice. Note that short HoG is more robust than a single-round gradients by reducing false alarms.

On the other hand, additive-noise attackers and unreliable clients (with low-quality data) would have similar short HoGs, but considered collectively, would be apart from other clients. Therefore, after excluding the sign-flipping attackers, we use a clustering method based on short HoG to distinguish the above two types of clients from other clients. Empirically, we choose DBSCAN~\cite{schubert2017dbscan} as the clustering method because it conforms to our intuition and yields the best results. Between these two types, additive-noise attackers tend to be {\em farther} away from other clients than unreliable clients as the attackers add deliberate perturbations; nevertheless, a separation boundary could be learned by finding the largest gap over Euclidean distances. We also note that this is not a clear-cut line and further processing is needed which we discuss below in Step 3. The above intuition is depicted in Fig.~\ref{signflip}b.

\noindent \textbf{2) Targeted attack.} Targeted attackers intend to manipulate the global model toward a specific convergence point (e.g., misclassifying all dogs to cats). Such intention can be captured by our long HoG which {\em reinforces} their adversarial goal over the entire history and is also robust to short-term noises and {\em camouflage cases} in which some attackers may strategically behave benignly in some of the rounds in order to evade detection.
In MUD-HoG, we use K-means clustering with $K$=2 over long HoG to separate out targeted attackers, {\em after} excluding untargeted attackers detected in Step 1.

\vspace{5pt}
\noindent \textbf{3) Unreliable clients.} Finally, MUD-HoG identifies and separates unreliable clients from normal ones. After excluding all the detected malicious clients (targeted and untargeted), the unreliable clients become farther from the {\em median} client in terms of their {\em short HoG}. Rather than using clustering, in this case we find that the {\em cosine distance} is the most effective to detect them and hence adopt it in MUD-HOG.

\vspace{-0.1in}
\subsection{Detection of Malicious Clients}
\vspace{-0.1cm}
Based on the basic ideas discussed above, now we present all the technical details of how MUD-HoG detects different types of clients.
The server starts detection from round $\tau_0$ ($\tau_0=l=3$ in our experiments). 
 
\vspace{-0.1in}
\subsubsection{Detecting untargeted attackers using short HoG.}\label{sec:det_untar}
MUD-HoG first computes the {\em median} short HoG over all clients, as $\nabla_{med}^{sHoG} = median\{\nabla^{sHoG}_i \  | 1 \leq i \leq N\}$. Then, it flags a client $c_i$ as a sign-flipping attacker if
\begin{align}\label{dist}
 d_{cos}\left (\nabla_{med}^{sHoG}, \nabla^{sHoG}_i \right ) < 0,
\end{align} 
where the function $d_{cos}(\cdot)$ computes the cosine distance.
We note that an existing algorithm CONTRA~\cite{awan2021contra} also employs cosine distance to separate out {\em targeted attackers}. CONTRA computes the pair-wise distances between the gradients of all the clients, which therefore leads to a complexity of $O(N^2)$; in contrast, MUD-HoG uses median and thus the complexity is linear, $O(N)$, which is worth noting because FL often deals with a massive number of clients (e.g., IoT devices). Moreover, CONTRA does not handle unreliable clients.

Next, MUD-HoG proceeds to detecting additive-noise attackers after excluding the above detected sign-flipping attackers. We apply DBSCAN clustering on the short HoGs of all the remaining clients and obtain two groups -- (i) a smaller group ($g_l$) consisting of the additive-noise attackers and unreliable clients and (ii) a larger group ($g_h$) consisting of the rest of the clients.
Based on our above analysis that the additive-noise attackers are relatively farther from normal clients than unreliable clients (Fig.~\ref{signflip}b), MUD-HoG attempts to learn a separation boundary as follows. Recalculate $\nabla_{med}^{sHoG}$ as the median short HoG of group $g_h$, and construct $\mathbf{d}=\{d_{Euc}(\nabla_{med}^{sHoG}, \nabla_{i}^{sHoG})\}$ which is a set of Euclidean distances (denoted by $d_{Euc}(\cdot)$) between $\nabla_{med}^{sHoG}$ and each client $c_i \in g_l$. 
The reason we use Euclidean distance rather than cosine distance is that the former produces a larger separation over {\em unnormalized} short HoG (which we intend).
Then, we find the largest gap between any two consecutive values in the {\em sorted} list of the set $\mathbf{d}$, and use the mid-point of this gap as the separation boundary $d_{\phi}$. Thus, a client $c_i \in g_l$ is an additive-noise attacker if
\begin{align}\label{additive}
d_{Euc}\left (\nabla_{med}^{sHoG}, \nabla^{sHoG}_i \right) > d_{\phi}
\end{align}
for $1 \leq i \leq |g_l|$. The remaining clients in $g_l$ and the set $g_h$ will be handled in the next step.
The above detection of untargeted attackers is summarized as the pseudo-code of Lines $6-16$ in Algorithm~\ref{mudhog}.

\vspace{-0.1in}
\subsubsection{Detecting targeted attackers using long HoG.} 
After excluding the detected untargeted attackers as above, we compute the long HoG for each of the remaining clients, denoted by $\nabla^{lHoG}_i$.
Then, we apply $K$-means clustering with $K=2$ on all the computed long HoGs to obtain two groups of clients: the smaller group will consist of the targeted attackers and the other (bigger) group of the normal clients, based on our assumption that normal clients constitute more than half of the entire population.
In Algorithm~\ref{mudhog}, Lines $17$-$18$ corresponds to the detection of targeted attackers. 

\SetAlFnt{\small}
\begin{algorithm}[h!]
 \caption{MUD-HoG}
\label{mudhog}
\KwIn {
Gradients from round 1 to $\tau$, for each client $c_i$, denoted by $\nabla_i=\{\nabla_{1,i}, \nabla_{2,i}, \cdots, \nabla_{\tau-1,i} \}$, $i=1...N$. (Note that the server only keeps the latest $l$ gradient vectors and the {\em sum} of all $\tau-1$ gradients.)}
\KwOut{Normal clients $(\mathcal{C}_{norm})$, targeted attackers $(\mathcal{C}_{tar})$, untargeted attackers $(\mathcal{C}_{untar})$, and unreliable clients $(\mathcal{C}_{unrl})$}
Initialize $\mathcal{C}_{norm}, \mathcal{C}_{tar},  \mathcal{C}_{untar} = \emptyset, \mathcal{C}_{all} = \{c_i\}, 1 \leq i \leq N$\\
\For{\textit{round} $\tau=1$ to $\tau_0$}{ 
Aggregate gradients of all clients
}
\For{\textit{round} $\tau=\tau_0+1$ to $\mathcal{T}$}{
Compute short HoG $\nabla_i^{sHoG}$ and long HoG $\nabla_i^{lHoG}$ for each client $c_i$ \\
\medskip
\tcc{Detecting untargeted attackers}
Computer median short HoG $\nabla_{med}^{sHoG}$ over all $N$ clients \\
\For{$i=1$ to $N$}{
\If{\eqref{dist} holds}{
$\mathcal{C}_{untar} = \mathcal{C}_{untar} \cup \{c_i\}$ \tcp*{Sign-flipping attackers}
}
}
Apply DBSCAN clustering on short HoGs of $\mathcal{C}_{all} \setminus \mathcal{C}_{untar}$ to obtain two groups $g_l$ and $g_h$ \\
Compute $\nabla_{med}^{sHoG}$ of the larger group $g_h$\\
Compute $d_{Euc}$ between $\nabla_{med}^{sHoG}$ and each $\nabla_i^{sHoG}$ of the smaller group $g_l$\\
Find the separation boundary $d_{\phi}$ per \sref{sec:det_untar} \\
\For{$i=1$ to $N$ and $c_i \notin \mathcal{C}_{untar}$}{
\If{\eqref{additive} holds}{
$\mathcal{C}_{untar} = \mathcal{C}_{untar} \cup \{c_i\}$ \tcp*{Additive-noise attackers}
}
}
\medskip
\tcc{Detecting targeted attackers}
Apply $K$-means clustering with $K=2$ on long HoGs of $\mathcal{C}_{all} \setminus \mathcal{C}_{untar}$\\
$\mathcal{C}_{tar} = $ clients who belong to the smaller cluster\\
\medskip
\tcc{Detecting unreliable clients}
Recompute $\nabla_{med}^{sHoG}$ over $\mathcal{C}_{all} \setminus \{\mathcal{C}_{tar} \cup \mathcal{C}_{untar}\}$\\
\smallskip
Compute $d_{cos}$ between $\nabla_{med}^{sHoG}$ and each $\nabla_i^{sHoG}$ of $\mathcal{C}_{all} \setminus \{\mathcal{C}_{tar} \cup \mathcal{C}_{untar}\}$  \\
Recompute the separation boundary $d_{\phi}$  per \sref{sec:det_unrl} \\
\For{$i=1$ to $N$ and $c_i \notin \{\mathcal{C}_{tar} \cup \mathcal{C}_{untar}\}$}{
\If{\eqref{unrel2} holds}{
$\mathcal{C}_{unrl} = \mathcal{C}_{unrl} \cup \{c_i\} $
}
}
$\mathcal{C}_{norm} = \mathcal{C}_{all} \setminus \{\mathcal{C}_{tar} \cup \mathcal{C}_{untar} \cup \mathcal{C}_{unrl}\}$\\
\medskip
\tcc{Aggregate gradients over $\mathcal{C}_{norm}$ and $\mathcal{C}_{unrl}$}
$\boldsymbol{\nabla}_\tau = \sum_{i\in \mathcal{C}_{norm}} \frac{|\mathcal{D}_i|}{|\mathcal{D}|} \nabla_{\tau,i} + \alpha \sum_{i\in \mathcal{C}_{unrl}} \frac{|\mathcal{D}_i|}{|\mathcal{D}|} \nabla_{\tau,i}$\\
Update global model as
$\boldsymbol{w}_{\tau+1} = \boldsymbol{w}_{\tau} - \eta \boldsymbol{\nabla}_\tau$ \\
Send $\boldsymbol{w}_{\tau+1}$ back to all clients 
}
\Return $\mathcal{C}_{norm}, \mathcal{C}_{tar}, \mathcal{C}_{untar},  \mathcal{C}_{unrl}$
\end{algorithm}

\vspace{-0.1in}
\subsection{Detection of Unreliable Clients}\label{sec:det_unrl}
\vspace{-0.1in}

We are now left with a mixture of unreliable and normal clients. 
To distinguish them, MUD-HoG finds a new separation boundary $d_{\phi}$ as follows. Let $N{'}$ be the number of remaining clients and $\nabla_{med}^{sHoG}$ be the (updated) median short HoG of them. Let $\mathbf{d}'=\{d_{cos}(\nabla_{med}^{sHoG}, \nabla_{i}^{sHoG})\}$ be a set of {\em cosine} distances between $\nabla_{med}^{sHoG}$ and each client $c_j$ for $1 \leq i \leq N{'}$. The separation boundary $d_{\phi}$ is then determined from $\mathbf{d}'$ similarly as the above detection of additive-noise attackers (but here we use cosine distance). Then, a client $c_i$ is deemed unreliable if it satisfies the condition 
\begin{align}\label{unrel2}
d_{cos}\left (\nabla_{med}^{sHoG}, \nabla^{sHoG}_i \right) < d_{\phi}.
\end{align}
Note that the cosine distance is smaller when the angle between two vectors is larger, and that is why the condition `$<$' used in \eqref{unrel2} is opposite to that in \eqref{additive}. The unreliable clients are detected at Lines $19$-$24$ in Algorithm~\ref{mudhog} after exclusion of all types of attackers. 

Thus finally (in each FL round), MUD-HoG obtains the set of normal clients $\mathcal{C}_{norm}$ and the set of unreliable clients $\mathcal{C}_{unrl}$, after filtering out
$\mathcal{C}_{tar}$ and $\mathcal{C}_{untar}$. It then aggregates the gradients of normal and unreliable clients using \eqref{eq:agg_unrl} (or see Line $26$ in Algorithm~\ref{mudhog}), where unreliable clients are downscaled, and then updates the global model as
$\boldsymbol{w}_{\tau+1} = \boldsymbol{w}_{\tau} - \eta \boldsymbol{\nabla}_\tau$.
Clearly, since the gradients of malicious clients have been discarded, their negative impact is eradicated from the global model.

\vspace{-0.2cm}
\section{Performance Evaluation}\label{sec:ExpEvaluation}
\vspace{-0.1in}
In this section, we evaluate MUD-HoG in comparison with six state-of-the-art methods on two real datasets with various type of attacks.
\vspace{-0.15in}
\subsection{Experiment Setup}
\label{subsec:ExpSetup}
\vspace{-0.06in}
We consider a classification task on two datasets: (i)     MNIST~\cite{lecun1998mnist}: Our FL task is to train a deep model with 2 convolutional neural networks (CNN) followed by 3 fully connected layers \footnote{Adopt the model from \href{https://pytorch.org/tutorials/beginner/blitz/neural_networks_tutorial.html}{PyTorch tutorial}.}   to classify 10 digits.
    (ii) Fashion-MNIST~\cite{fashionMNIST}: We build a deep model with 6 CNN layers followed by two fully connected layers to classify 10 fashion classes. 

\vspace{3pt}
\noindent \textbf{Hyper-parameters.} We train the FL model with SGD optimizer (learning rate = 1e-2, momentum = 0.5 for MNIST and 0.9 for Fashion-MNIST, and weight-decay = 1e-4 for Fashion-MNIST) over 40 communication rounds, 4 local epochs; other setup details are similar to~\cite{Wan2021RobustFL}. 
We use the window size of $l=3$ for calculating the moving average short HoG. Our algorithm triggers only after $\tau_0=3$ rounds to accumulate enough HoGs. Since, the server stores only $l+1$ gradient vectors ($l$ latest and a {\em sum} of all previous vectors) to compute HoGs, it never runs into storage related issues. 
Moreover, we make a firm decision about malicious clients if they are detected in two consecutive rounds. 
Therefore, our algorithm can only detect malicious clients at least after $4$ rounds. 

To simulate non-IID data, we divide the datasets into 40 clients as disjoint portions that follows {\em Dirichlet distribution} with hyperparameter 0.9, as also adopted by \cite{Wan2021RobustFL, bagdasaryan2020backdoor}.
Besides normal clients, our FL system consists of unreliable clients (up to 10\% of total clients), and malicious clients (up to 47.5\% of total clients), as detailed below.

\vspace{3pt}
\noindent{\bf Untargeted attacks.}
     (i) \textit{Sign-flipping (SF)} --  We flip the sign of gradients of the malicious clients without enlarging the magnitudes in our FL setup, which makes the detection more challenging.
    (ii) \textit{Additive-noise (AN)} -- We add a Gaussian noise with $\mu=0$ and $\sigma=0.01$ to the gradients of attackers. 
    
\vspace{3pt}
\noindent{\bf Targeted attacks.}
    (i) \textit{Label-flipping (LF)} --Before training the local model, attacker flips label of digit "1" to "7" in its local MNIST dataset, and label ("1-Trouser") to ("7-Sneaker") in Fashion-MNIST dataset. 
    (ii) \textit{Multi-label-flipping (MLF)} -- Attacker flips the labels of few source classes to a targeted class in its local dataset. For MNIST and Fashion-MNIST (in brackets) datasets, we flip three source labels of digits "1" ("1-Trouser"), "2" ("2-Pullover"), and "3" ("3-Dress") to a target label "7" ("7-Sneaker").

\vspace{3pt}
\noindent{\bf Unreliable clients.} We simulate them to mimic a real-life scenario of low-end smartphone with poor-resolution camera and computing power. We use {\em Gaussian smoothing} (kernel size$=7$, $\sigma=50$) to blur $50\%$ of the local image dataset; and simulate low computing power by training over randomly selected portion of 30\% of local dataset. We set $\alpha=0.5$ to downscale the unreliable clients. 

To simulate heterogeneous FL scenarios, we consider two different series of experiments with upto $47.5\%$ malicious clients (including untargeted and targeted attack) and upto $10\%$ unreliable clients. 
We configure 12 different experimental setups with increasing numbers of unreliable and malicious clients as follows.

\begin{itemize}
\vspace{-0.05in}
    \item \textit{Series of Exp1} consists of $a=\min\{i,4\}$ unreliable clients, $b=\min\{i,6\}$ additive-noise  attackers, $c=\min\{i,5\}$ sign-flipping attackers, $d=(i+2)$ label-flipping attackers, and $(40-a-b-c-d)$ normal clients; where $i=\{1,2,3,4,5,6\}$. 

    \item \textit{Series of Exp2} consists of $a=\min\{i,4\}$ unreliable clients, $b=\min\{i,6\}$ additive-noise attackers, $c=\min\{i,5\}$ sign-flipping attackers, $d=(i+2)$ multi-label-flipping attackers, and $(40-a-b-c-d)$ normal clients; where $i=\{1,2,3,4,5,6\}$. 
\end{itemize}

\noindent \textbf{Evaluation metrics.}
The performance of MUD-HoG is measured in terms of precision, recall, accuracy, and detection ratio. We define {\em detection ratio} ($r$) as 
\begin{equation}
r=\frac{\sum_{\tau=1}^{\mathcal{T}} \sum_{i \in \mathcal{C}_{x}} \mathbbm{1} (c_i \text{ detected at } \tau) } {\mathcal{T} \sum_x |\mathcal{C}_{x}| }
\label{eq:detection_coverage}
\end{equation}
where $\mathcal{C}_x$ is either $\mathcal{C}_{tar}$, $\mathcal{C}_{untar}$ or $\mathcal{C}_{unrl}$, and not all of them are empty.
The higher the detection ratio (closer to $100\%$), the better algorithm is.

\vspace{5pt}
\noindent \textbf{Benchmark algorithms.}
In addition to FedAvg \cite{mcmahan2017communication}, a popular algorithm in FL, we compare our proposed MUD-HoG algorithm with five other algorithms. 
They are: (i) coordinate-wise Median (or Median for short)~\cite{yin2018byzantine}, (ii) GeoMed~\cite{chen2017distributed}, (iii) Krum~\cite{blanchard2017machine}, (iv) Multi-krum (or MKrum for short)~\cite{blanchard2017machine}, and (v) FoolsGold~\cite{fung2020limitations}. We borrowed the source code of these existing algorithms from~\cite{Wan2021RobustFL}.

\vspace{-0.05in}
\subsection{Experimental Results}
\vspace{-0.05in}
\subsubsection{Overall performance.}
\begin{figure*}[tb]
    \centering
    \subfloat[Series of Exp1]{
        \includegraphics[width=0.24\linewidth]{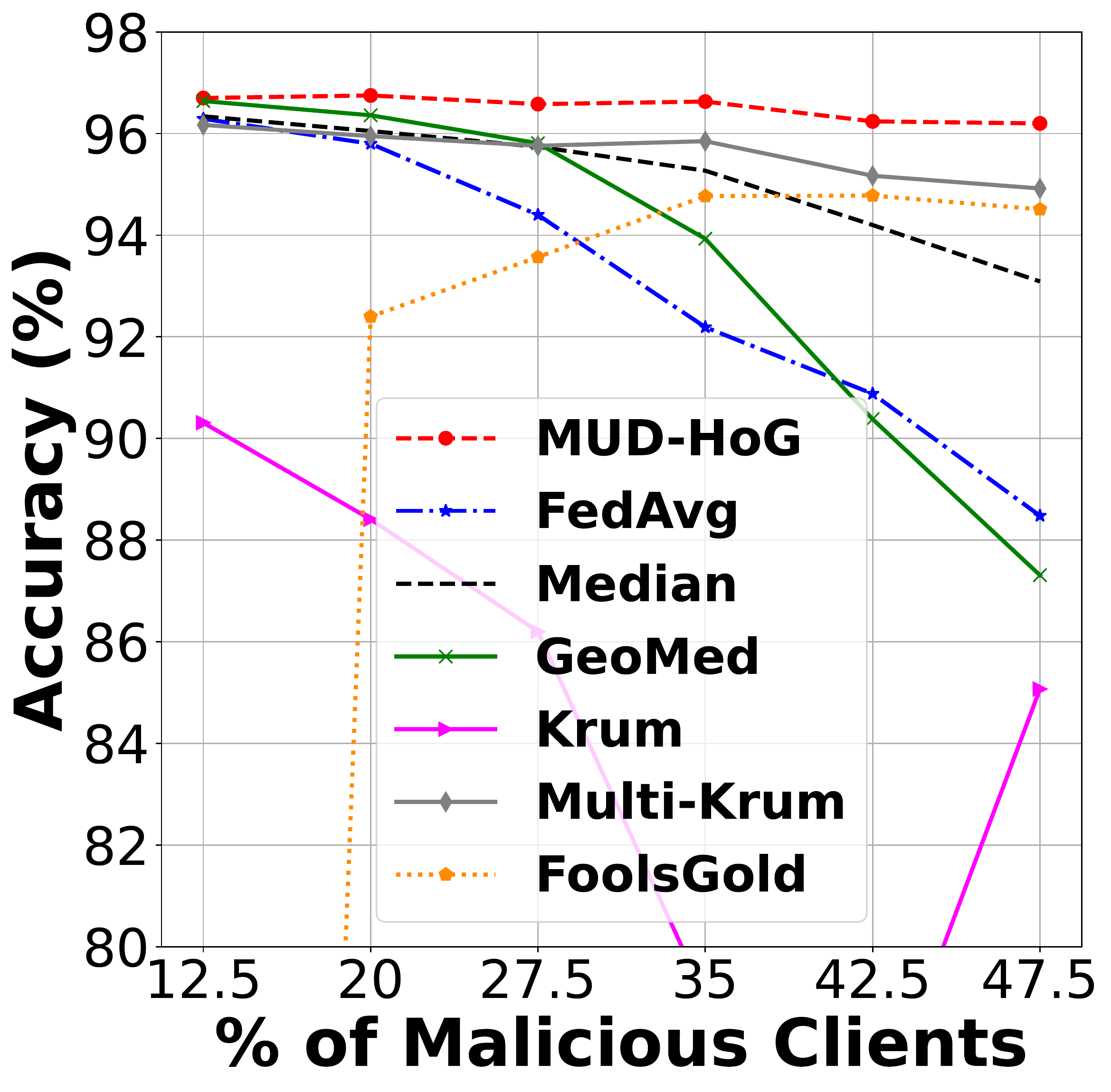}%
        \label{fig:acc_sExp1_MNIST}
    }
    \subfloat[Series of Exp2]{
        \includegraphics[width=0.24\linewidth]{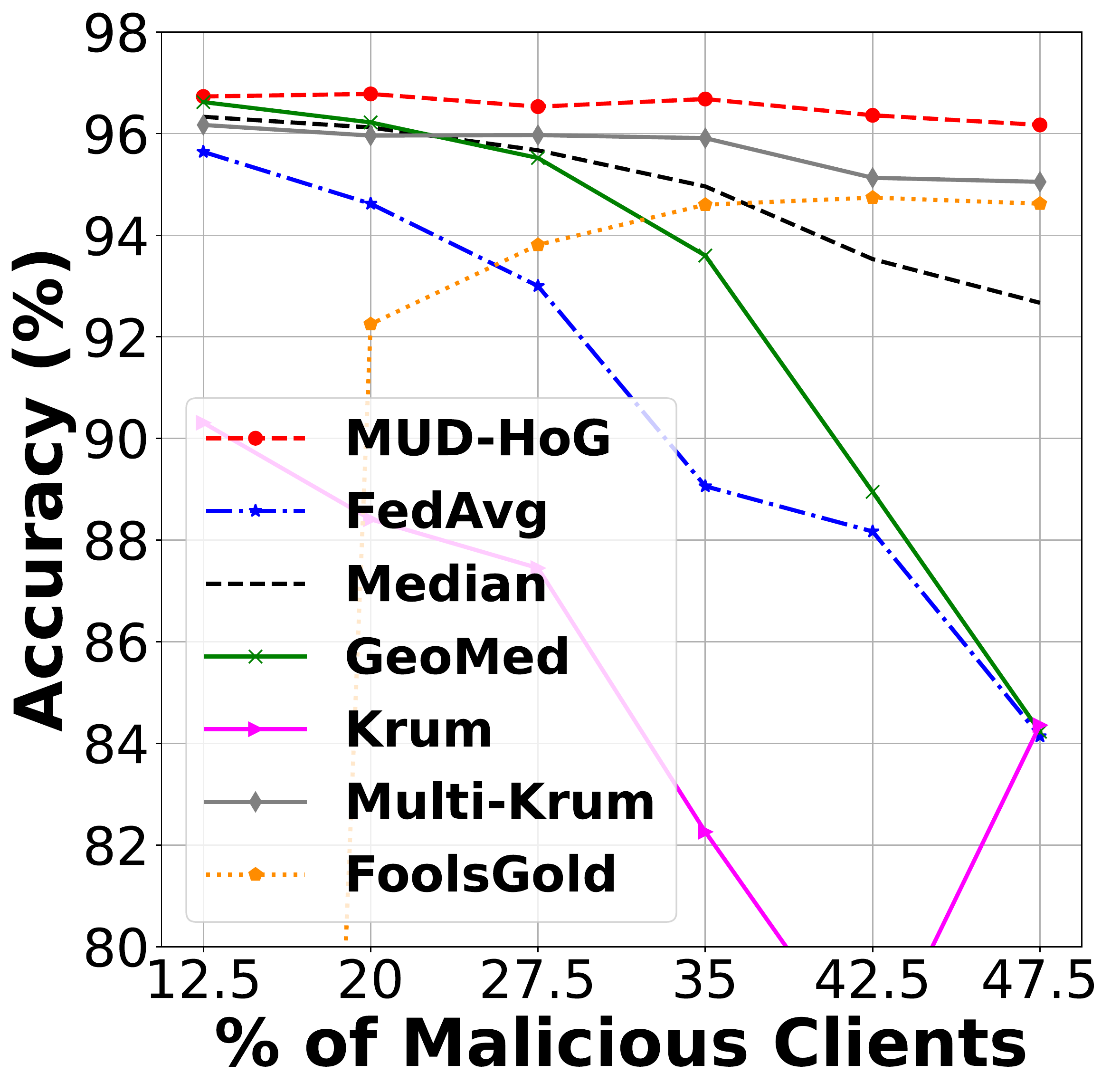}%
        \label{fig:acc_sExp2_MNIST}
    }
        \subfloat[Series of Exp1]{
        \includegraphics[width=0.24\linewidth]{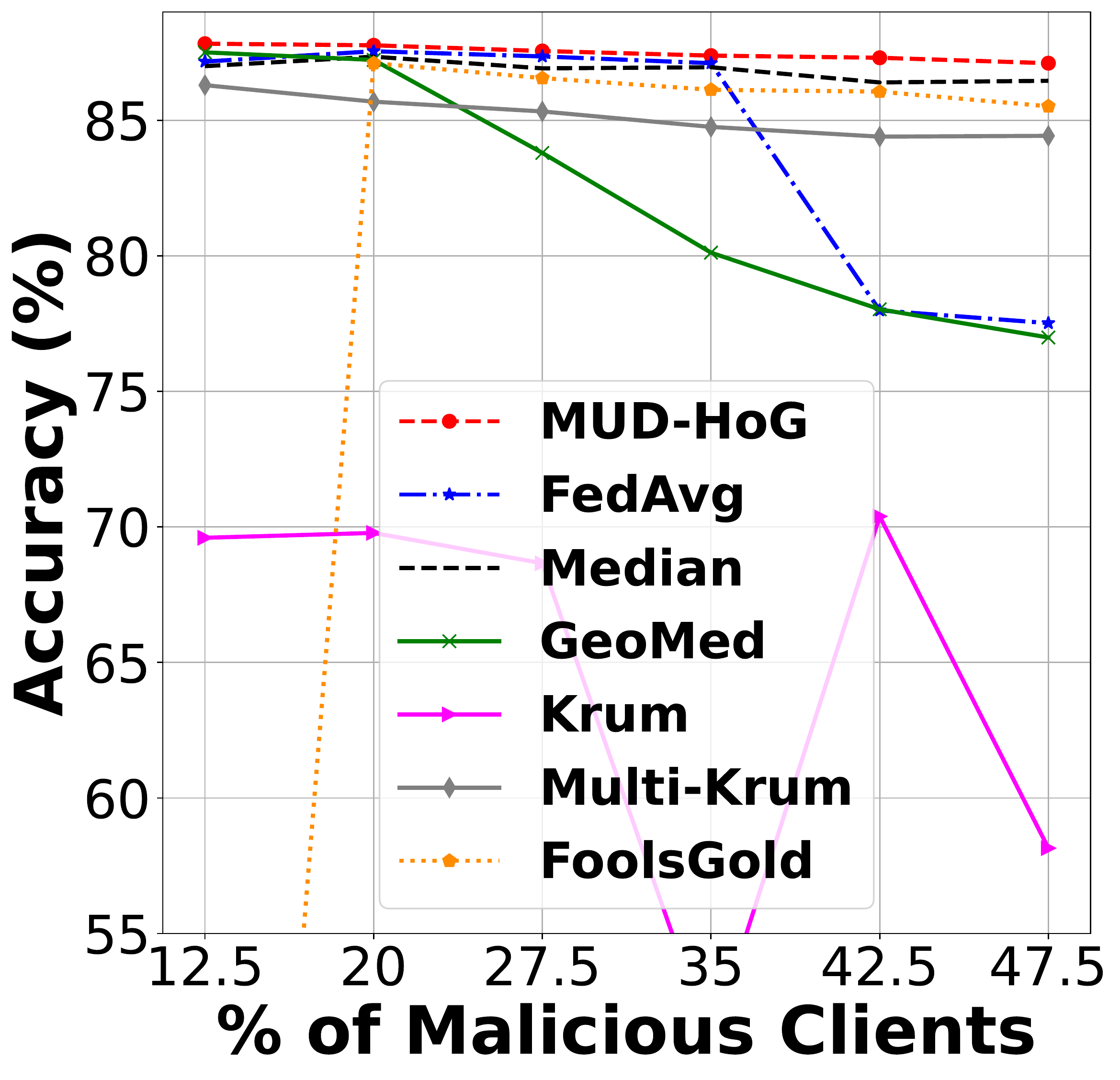}%
        \label{fig:acc_sExp1_FashionMNIST}
    }
    \subfloat[Series of Exp2]{
        \includegraphics[width=0.24\linewidth]{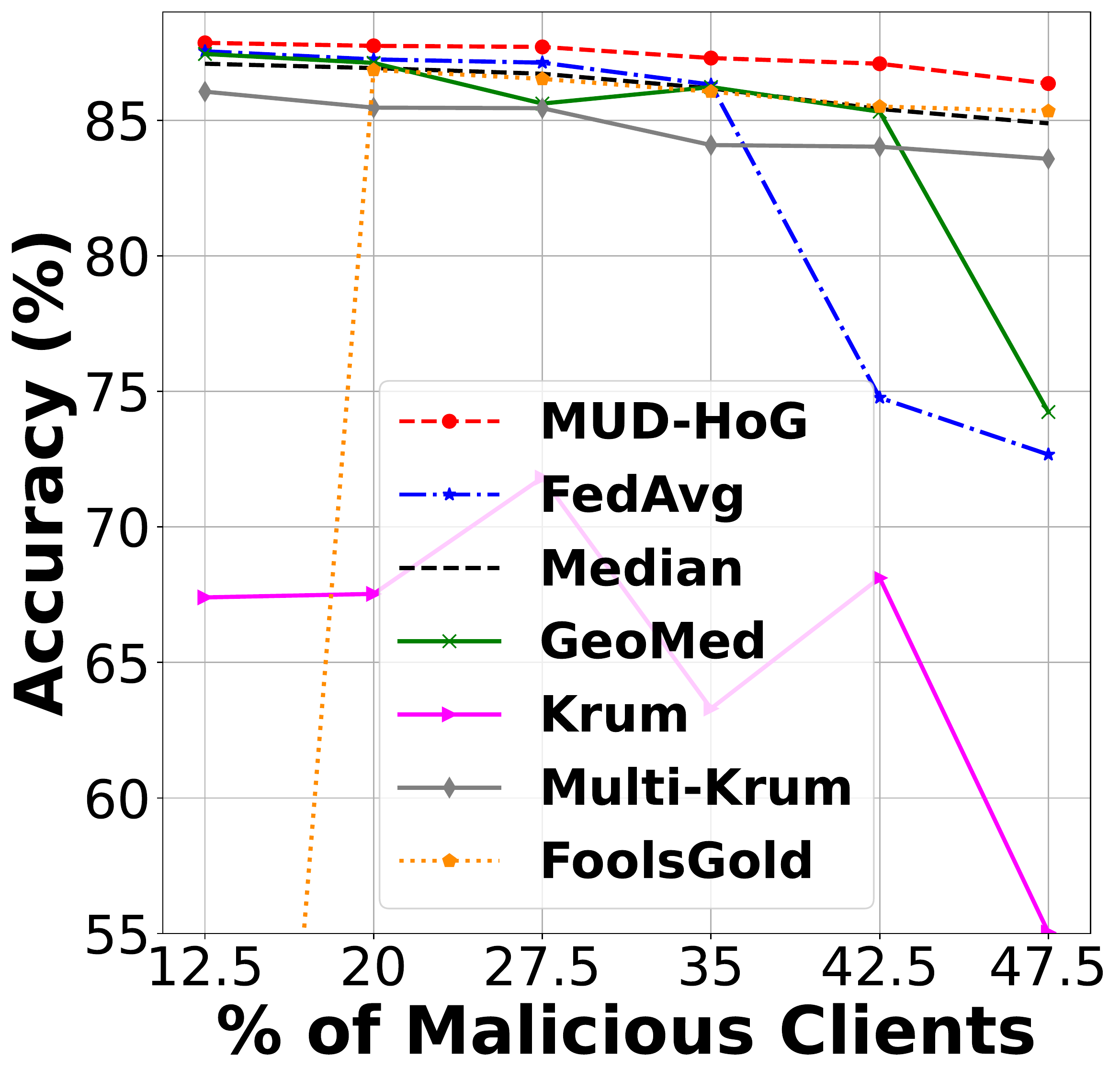}%
        \label{fig:acc_sExp2_FashionMNIST}
    }

    \caption{Accuracy vs. the percentage of malicious clients. (a) and (b) are results on the \textit{MNIST} dataset. (c) and (d) are results on the \textit{Fashion-MNIST} dataset.}
     \label{fig:acc_MNIST}
     \vspace{-0.2in}
\end{figure*}

Fig. \ref{fig:acc_MNIST} shows the accuracy of 12 setups for series of Exp1 and Exp2 for MNIST and Fashion-MNIST datasets under the above seven benchmark algorithms.
We observe that over all 12 setups with multiple types of attacks, MUD-HoG always achieves the best accuracy.

It is consistently observed that when increasing percentage of malicious clients from 12.5\% to 47.5\% of the total number of clients, Krum and FoolsGold show fluctuated performance and poor performance at a certain level of attacks, some other algorithms such as FedAvg, GeoMed, Median, and MKrum continuously drop their accuracy.
In contrast, our proposed MUD-HoG maintains robust performance against multiple levels of heterogeneous attacks. 

For \textbf{MNIST} dataset shown in parts (a) and (b) of Fig. \ref{fig:acc_MNIST}, initially GeoMed performs as good as MUD-HoG, but when the level of attacks are increased more than 35\%, GeoMed drops its accuracy by 9.33\% and 12.39\% while MUD-HoG only drops 0.5\% and 0.56\% in series of Exp1 and Exp2, respectively. 
When compared to the second-best algorithm, i.e. MKrum, the proposed algorithm gained upto 1.28\% and 1.12\% higher accuracy in series of Exp1 and Exp2, respectively.

For \textbf{Fashion-MNIST} dataset shown in parts (c) and (d) of Fig.~\ref{fig:acc_MNIST}, GeoMed achieves comparative results as MUD-HoG at a low level of attacks for both series; however, GeoMed drops performance significantly at the high level of attacks. 
For instance, in series of Exp1 and Exp2, while MUD-HoG's accuracy only drops by 0.72\% and 1.5\% (when increasing percentage of attacks from 12.5\% to 47.5\%), GeoMed's accuracy drops by 10.52\% and 13.21\%, respectively. 
When compared to the second-best algorithm, i.e., Median, MUD-HoG gains upto 0.65\% and 1.47\% accuracy in series of Exp1 and Exp2, respectively.

\vspace{-0.2in}
\subsubsection{Precision and Recall.}
\begin{figure}[h]
    \centering
    \subfloat[Precision of class "7"]{
        \includegraphics[width=0.24\linewidth]{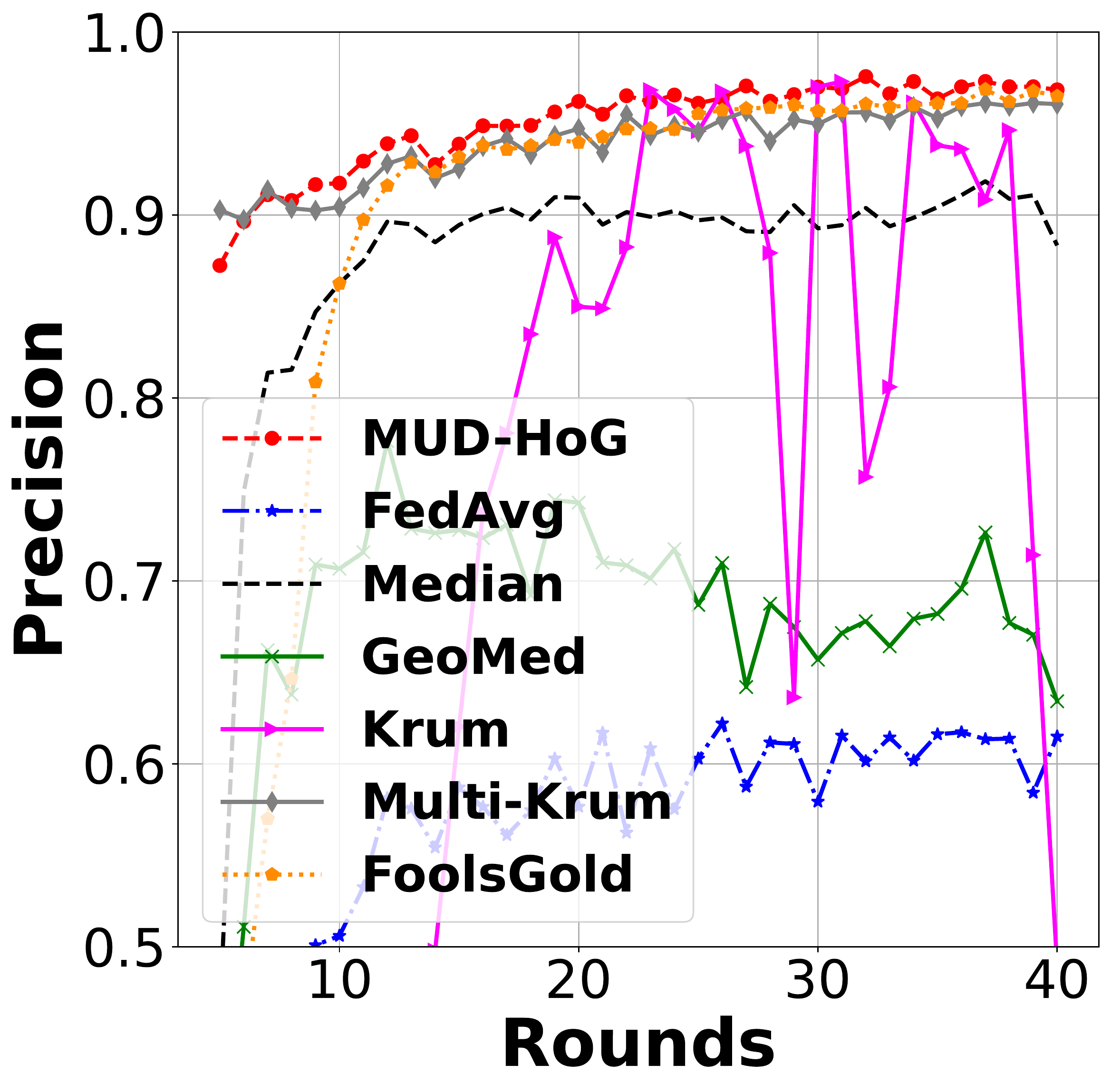}%
        \label{fig:precision_target7_sExp2_7MLF_MNIST}
    }
    \subfloat[Recall of class "2"]{
        \includegraphics[width=0.24\linewidth]{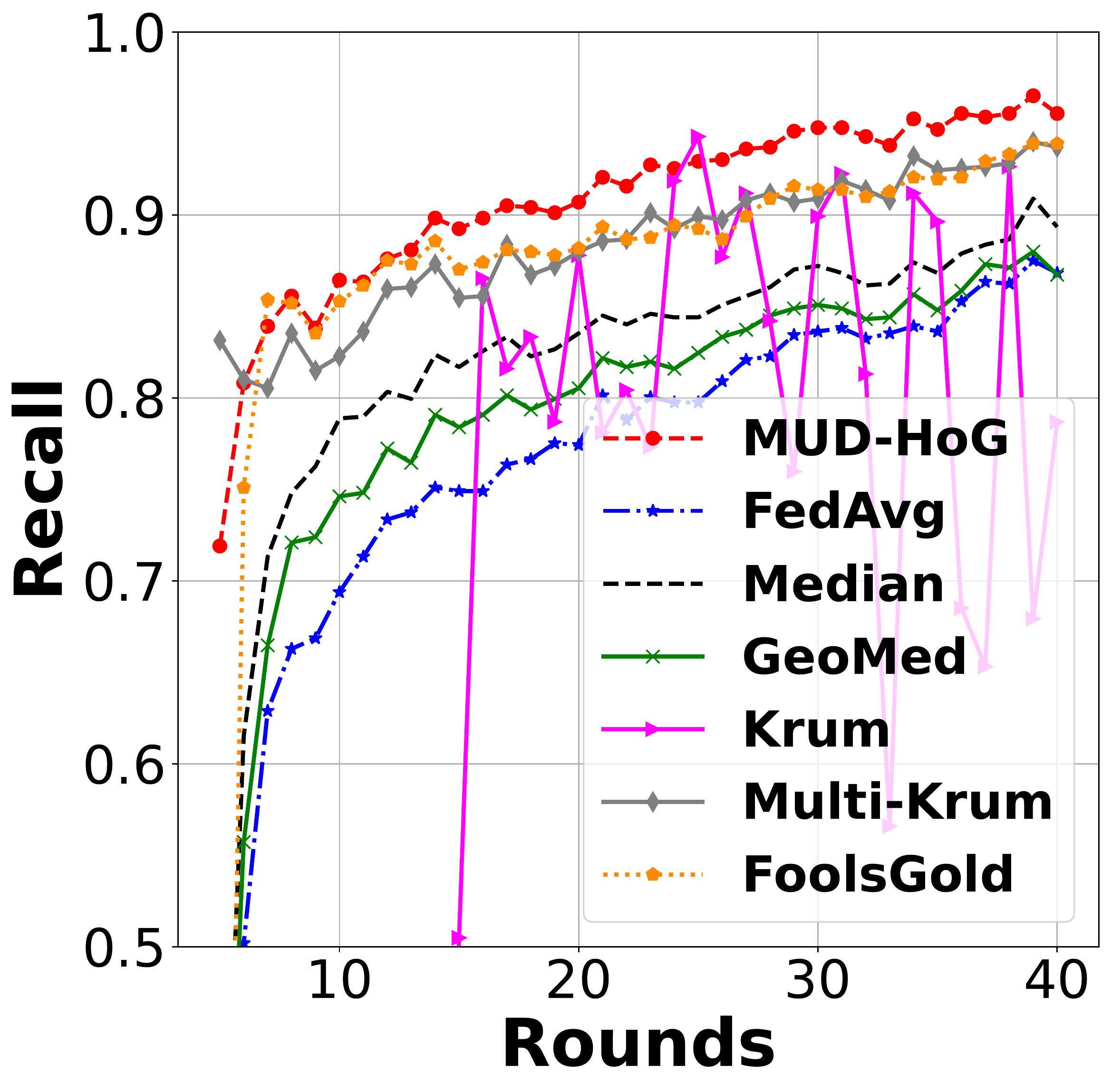}%
        \label{fig:recall_source2_sExp2_7MLF_MNIST}
    }
    \subfloat[Precision of class "7"]{
        \includegraphics[width=0.24\linewidth]{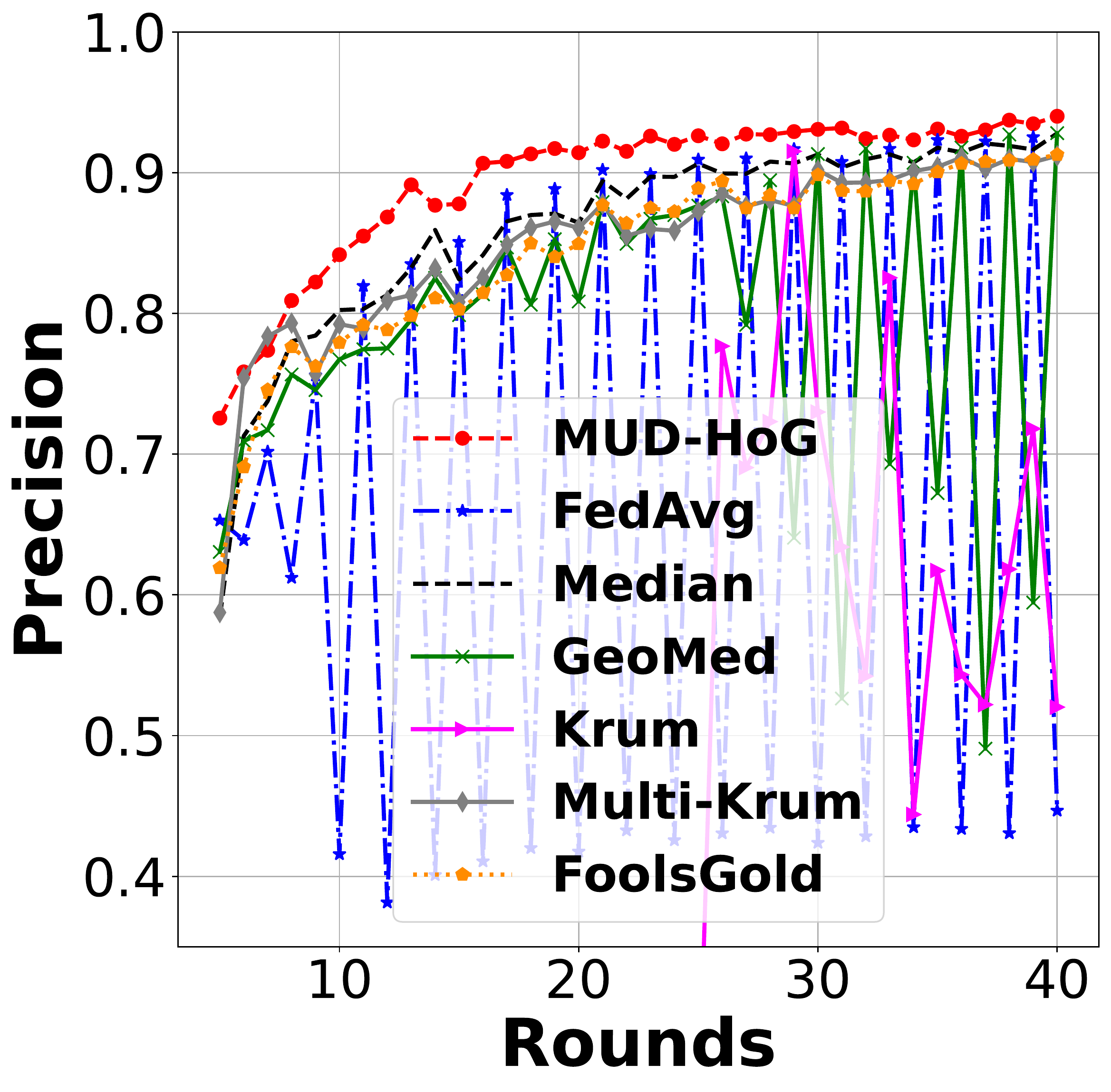}%
        \label{fig:precision_target7_sExp2_7MLF_FashioMNIST}
    }
    \subfloat[Recall of class "2"]{
        \includegraphics[width=0.24\linewidth]{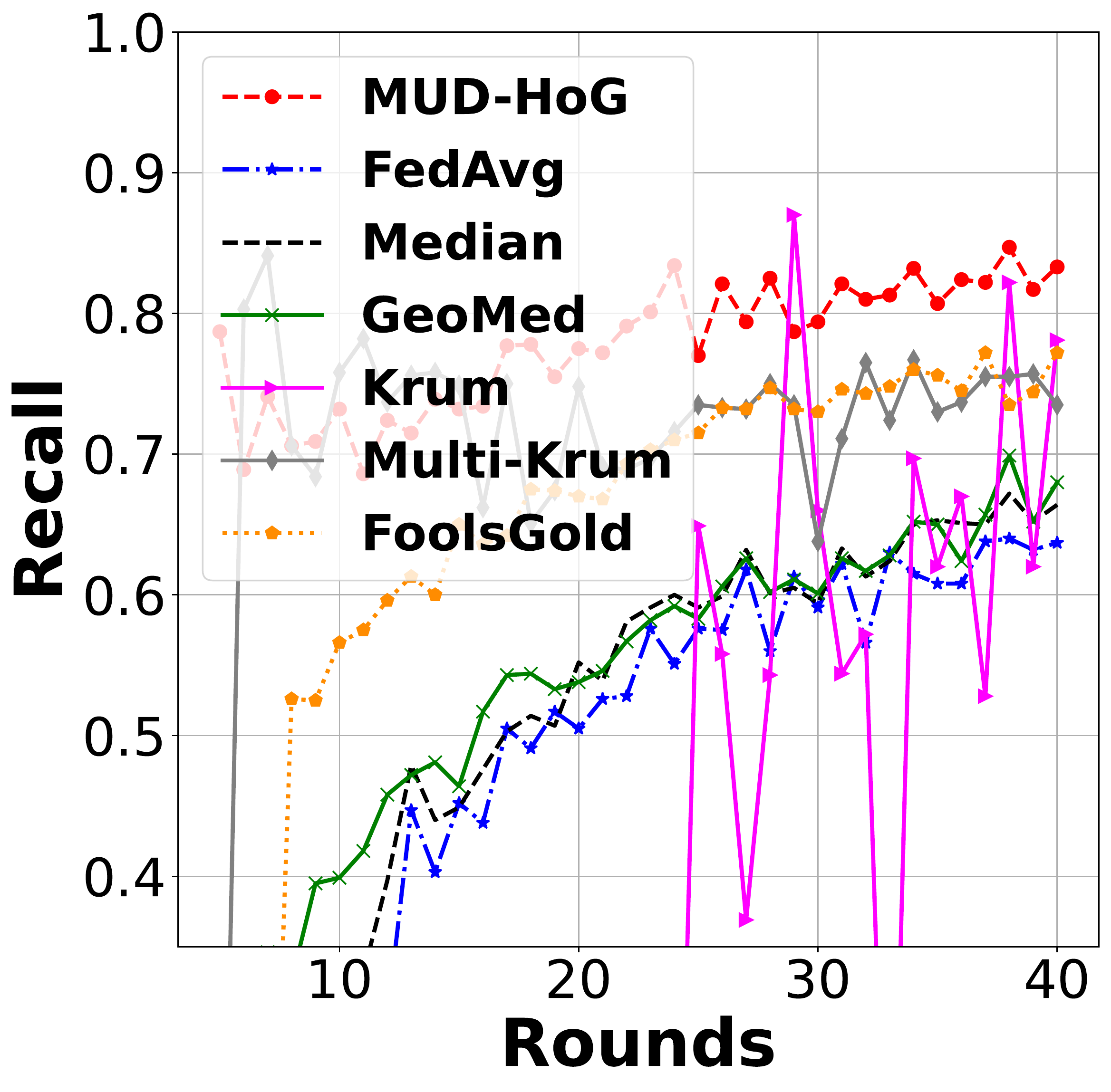}%
        \label{fig:recall_source2_sExp2_7MLF_FashionMNIST}
    }
    \caption{Results for {\it Series of Exp2} with 42.5\% malicious clients. "2" and "7" are the source and target classes, respectively. (a) and (b) are results on the \textit{MNIST} dataset. (c) and (d) are results on the \textit{Fashion-MNIST} dataset.}
     \label{fig:precision_target7_recall_source2_sExp2_7MLF_MNIST}
     \vspace{-0.2in}
\end{figure}

To make a fair comparison with other algorithms (i.e., Krum, MKrum, FoolsGold) that ware designed specifically for targeted attacks, we plot \textit{precision} of the targeted class (i.e., number of samples correctly classified as the targeted class over all samples predicted as the targeted class), and \textit{recall} of a source class (i.e., number of samples correctly classified as the source class over all ground-truth samples of the source class) for MNIST and Fashion-MNIST datasets in Fig.~\ref{fig:precision_target7_recall_source2_sExp2_7MLF_MNIST}. 
Here, FedAvg, GeoMed, Median or even Krum obtain poor performance and highly fluctuated precision of targeted class and recall of source class because they could not defend targeted attacks.
On the flip side, though MKrum and FoolsGold show quite good precision, their values are lower than MUD-HoG for both the datasets. 

\vspace{-5pt}
\subsubsection{Detection Ratio.}
We keep track of detected rounds for each type of clients during the course of FL training with MUD-HoG algorithm. 
Table~\ref{tab:DetectMalicious} reports detection ratio (defined in Eq.~\ref{eq:detection_coverage}) for each type of clients, and their first round of detection (presented inside brackets) for a setup in series of Exp1 and Exp2 with 27.5\% malicious clients. 
We observe that the sign-flipping and additive-noise attackers are detected immediately at round $4$, which is the earliest round when the MUD-HoG algorithm could provide a firm decision.

\begin{table}[h!]
\vspace{-0.2in}
    \caption{Detection ratio $r$ (\%) and the earliest round (\textbf{1\textsuperscript{st}rnd}) that detects the client type (round number in brackets), with 27.5\% malicious clients. [\textbf{SF}: Sign-flipping, \textbf{AN}: Additive-noise, \textbf{LF}: Label-flipping, \textbf{MLF}: Multi-label-flipping, \textbf{UR}: Unreliable]}
    \label{tab:DetectMalicious}
    \vspace{0.1in}
    \centering
    \resizebox{.7\textwidth}{!}{
    \begin{tabular}{l@{\hspace{0.6em}} l@{\hspace{0.65em}} | c@{\hspace{0.65em}} c@{\hspace{0.65em}} | c@{\hspace{0.65em}}  c@{\hspace{0.65em}} c@{\hspace{0.65em}} }
        \hline
 &      & \multicolumn{2}{c|}{\textbf{MNIST}}  & \multicolumn{2}{c}{\textbf{Fashion-MNIST}} \\ \hline
\textbf{Type} & \textbf{Detection}     & \textbf{Exp1} & \textbf{Exp2}  & \textbf{Exp1}  &  \textbf{Exp2} \\
        \hline
\textbf{SF} & \textbf{$r$ (1\textsuperscript{st}rnd)}        & 90.0 (4)         &  90.0 (4)              & 90.0 (4)              & 90.0 (4)  \\ \hline

 \textbf{AN} & \textbf{$r$ (1\textsuperscript{st}rnd)}       & 90.0 (4)         & 90.0 (4)               & 90.0 (4)             & 90.0 (4) \\ \hline

\textbf{LF}  & \textbf{$r$ (1\textsuperscript{st}rnd)}      & 87.5 (5)         &  -             & 85.0 (6)           & - \\ \hline

\textbf{MLF} & \textbf{$r$ (1\textsuperscript{st}rnd)}   & -        & 90.0 (4)               & -         & 85.0 (6) \\ \hline
\hline
\multicolumn{2}{l|}{\textbf{Overall rate} $r$ (\%)} & 88.9 &  90.0 & 87.7 & 87.7 \\ \hline

\textbf{UR} 
                    &\textbf{$r$ (1\textsuperscript{st}rnd)}               & 87.5 (5)        & 87.5 (5)                & 85.0 (6)            & 85.0 (6)  \\ \hline

\end{tabular}
}
\vspace{-0.1in}
\end{table}

For MNIST dataset, overall, we can detect all malicious clients at detection ratio (calculated over all types of clients) $88.9\%$ and $90.0\%$ for a setup in series of Exp1 and Exp2, respectively.
Since FL training is done over 40 rounds and the earliest detection round is 4, upper bound of detection ratio can be at most $90.0\%$.  
And we can see in Exp2 of MNIST, MUD-HoG can detect MLF at round $4$, which is as early as SF or AN, resulting in 90.0\% of detection ratio. 
Next, for Fashion-MNIST dataset, our algorithm detects targeted attacks (i.e., LF and MLF) a bit slower than the case in MNIST, but the overall detection ratio is still above 87\%. 
Finally, for unreliable clients (last two rows in Table~\ref{tab:DetectMalicious}), in all experiments, MUD-HoG achieves firm results of all unreliable clients from round $5$ and round $6$ for MNIST and Fashion-MNIST datasets, respectively. 
As a result, the detection ratio for unreliable clients is above 85.0\%. 

\vspace{-0.1in}
\subsection{Discussions and Limitations}
\vspace{-0.1in}

\vspace{3pt}
\noindent \textbf{Convergence analysis.} Based on our experimental results (see Fig.~\ref{fig:test_accuracy_loss_sExp4_7MLF_MNIST} and Fig.~\ref{fig:test_accuracy_loss_sExp4_7MLF_FashionMNIST}), the loss of the global model stabilizes in 40 FL rounds for both the datasets even in the presence of 42.5\% clients posing different types of attacks and having non-IID data. This indicates that MUD-HoG can achieve convergence in rather adversarial scenarios. Although the presence of malicious clients initially diverges the global model from its objective, excluding them from aggregation, as MUD-HoG did, rectifies the SGD process back to normal as defined in~\cite{nguyen2019new}. 
In future work, we plan to incorporate a rigorous theoretical analysis of convergence for our approach.

\vspace{3pt}
\noindent \textbf{More strategic attacks.} While we have experimentally shown that MUD-HoG is robust to various untargeted and targeted attacks in the presence of a large number of malicious clients, it may still miss out attackers who perform stealthy or highly strategic targeted attacks (some are formally defined in~\cite{chen2017distributed}). Besides, an attacker may implant a certain trigger pattern into some training/test data to inject corruption~\cite{wang2020attack, bhagoji2019analyzing}, known as {\em backdoors}. Such attacks are more evasive since they are only triggered when the particular pattern arises, while the overall performance is almost not affected. Currently, MUD-HoG has not been specifically designed to defend backdoor attacks but this would be an interesting direction to explore.

\vspace{-0.1in}
\section{Conclusion}\label{conclusion}
\vspace{-0.1in}
While federated learning (FL) offers a privacy-preserving framework for collaborative training of ML models, it is susceptible to adversarial attacks. This paper has proposed a new approach called MUD-HoG to detect malicious clients who launch untargeted or targeted attacks and unreliable clients who possess low-quality data, and offers a fine-grained classification of four types of participants. We introduce the concept of long-short HoG and select appropriate distance and similarity measures to identify different types of attacks and clients. MUD-HoG excludes malicious contributions but exploits unreliable clients' contributions to maximize the utility of the final global model. 
Experimental results confirm that MUD-HoG is robust against malicious and unreliable clients and produces a global model with higher accuracy than state-of-the-art baselines. It can detect a mixture of multiple types of attackers and unreliable clients in non-IID settings even when the ratio of attackers is close to half.
In future work, we plan to investigate more challenging and dynamic settings where attackers may vary attack types and clients may even switch roles (attackers, unreliable, normal, etc.) over time. More extensive experiments will also be conducted.

\vspace{10pt}
\noindent{\bf Acknowledgements:} This work is partially supported by the NSF grant award \#2008878 (FLINT: Robust Federated Learning for Internet of Things) and the NSF award \#2030624 (TAURUS: Towards a Unified Robust and Secure Data Driven Approach for Attack Detection in Smart Living).

\newpage

\makeatletter
\def\@seccntformat#1{\@ifundefined{#1@cntformat}%
   {\csname the#1\endcsname\quad}  
   {\csname #1@cntformat\endcsname}
}
\let\oldappendix\appendix
\renewcommand\appendix{
    \oldappendix
    \newcommand{\section@cntformat}{\appendixname~\thesection\quad}
}
\makeatother

\appendix

\section{Additional Experimental Results}
\subsection{Performance improvement over rounds}
We consider a specific setup with 42.5\% malicious clients, for both the datasets to evaluate the improvement of the accuracy of all the algorithms over FL rounds. 

\begin{figure}[h]
\centering

    \subfloat[Accuracy]{
        \includegraphics[width=0.30\linewidth]{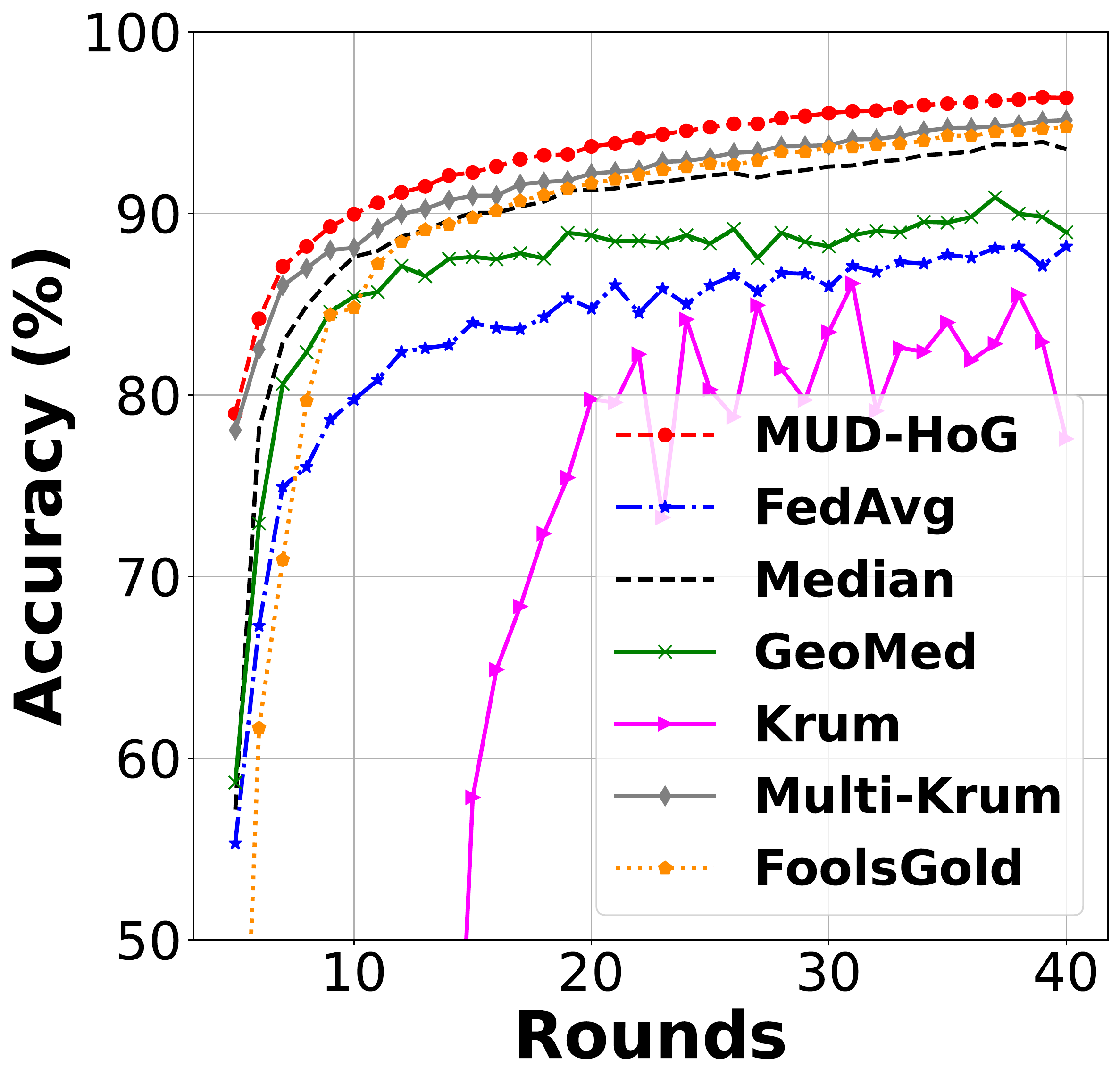}%

    }
    \hspace{0.2in}
    \subfloat[Loss]{
        \includegraphics[width=0.3\linewidth]{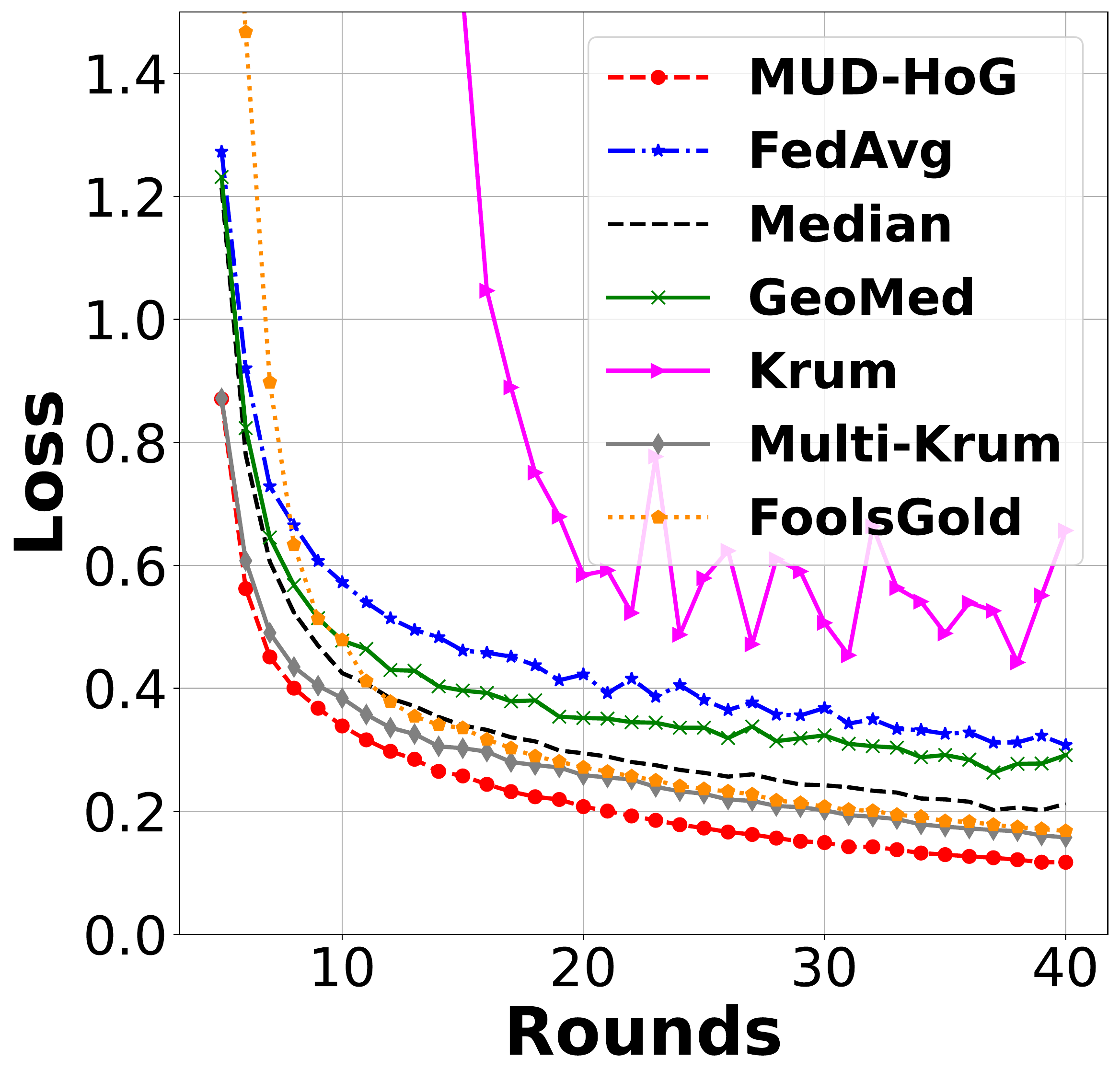}%
    }
    
\caption{Performance improvement of global model on \textit{MNIST} in {\it Series of Exp2} with 42.5\% malicious clients.}
\label{fig:test_accuracy_loss_sExp4_7MLF_MNIST}
\vspace{-0.2in}
\end{figure}

We plot test accuracy and loss from round 5 to the final round 40 for MNIST dataset in Fig.~\ref{fig:test_accuracy_loss_sExp4_7MLF_MNIST} using global model. 
It is obvious to see that MUD-HoG obtains an upper bound of test accuracy and an lower bound of test loss over the course of FL training.
While some algorithms show fluctuated performance during training such as Krum with a high fluctuation, or FedAvg and GeoMed with smaller fluctuations,
the other state-of-the-art algorithms designed against attackers such as Median, MKrum, FoolsGold and MUD-HoG show smooth improvement as training progresses.
Among these algorithms, we also observe in Fig.~\ref{fig:test_accuracy_loss_sExp4_7MLF_MNIST} that the gap of test loss between MUD-HoG and the second-best algorithm is increasing over the course of FL training.

\begin{figure}[h]
\centering

    \subfloat[Accuracy]{
        \includegraphics[width=0.30\linewidth]{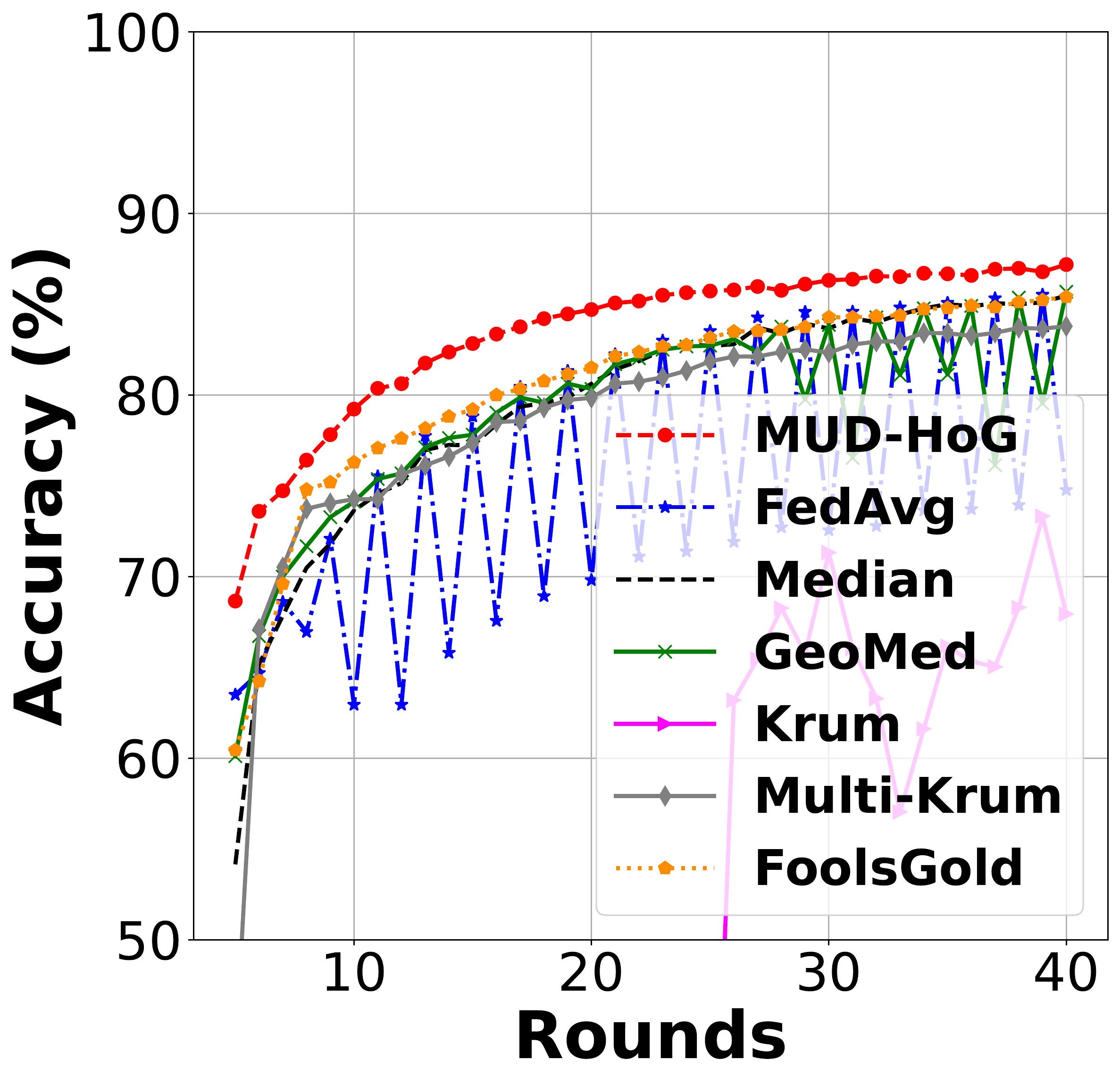}
    }
    \hspace{0.2in}
    \subfloat[Loss]{
        \includegraphics[width=0.3\linewidth]{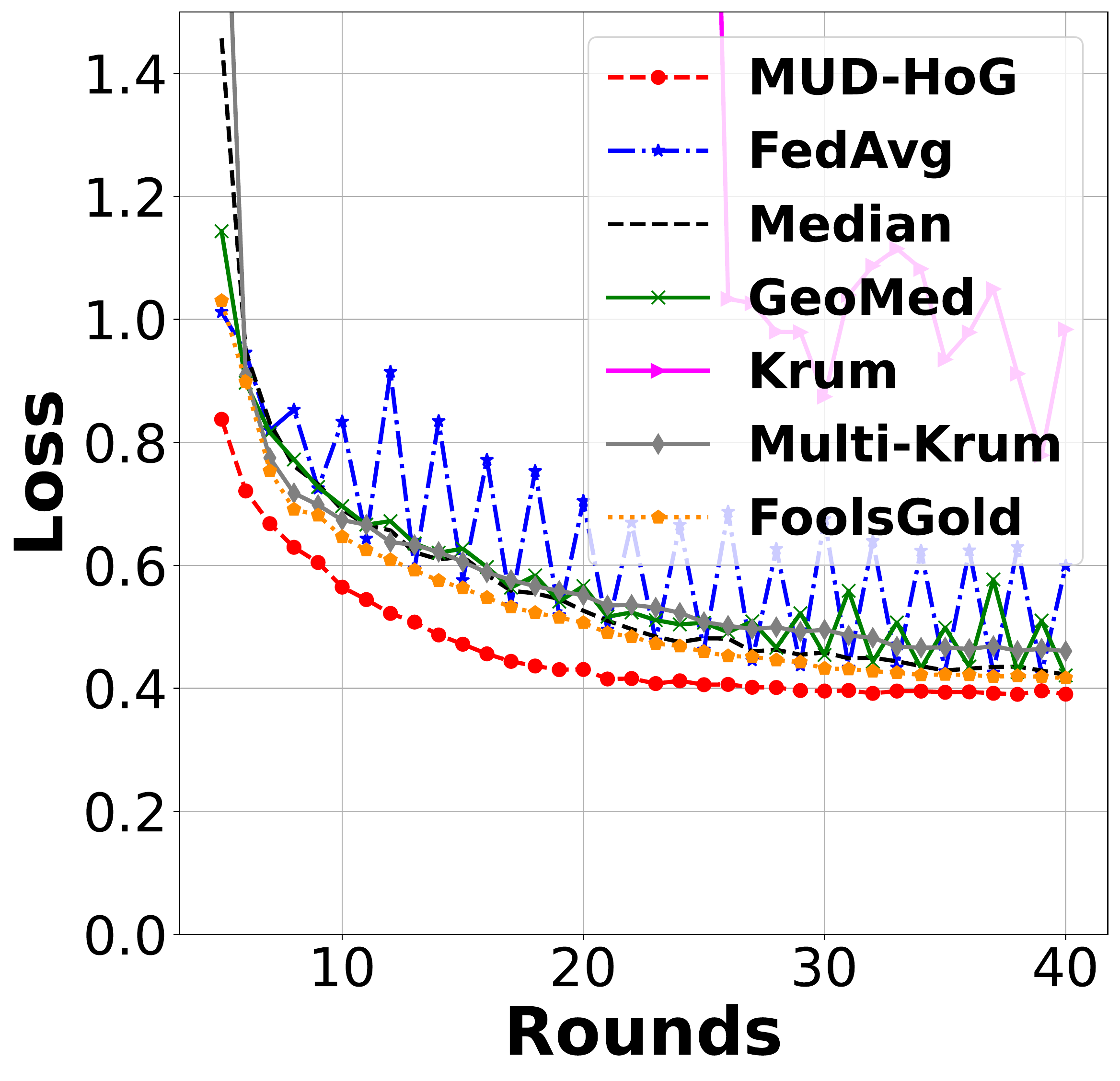}
}

\caption{Performance improvement of global model on \textit{Fashion-MNIST} in {\it Series of Exp2} with 42.5\% malicious clients.}
\label{fig:test_accuracy_loss_sExp4_7MLF_FashionMNIST}
\end{figure}

Fig.~\ref{fig:test_accuracy_loss_sExp4_7MLF_FashionMNIST} shows test accuracy and loss for Fashion-MNIST dataset. 
Similar to MNIST's results, we can see that among all evaluated algorithms, MUD-HoG obtains the highest accuracy and the lowest loss for all training rounds. 
The fluctuation of FedAvg and GeoMed is more severe with high variance, so the final accuracy of these algorithms are not really reliable. This is the reason why FedAvg and GeoMed can obtain accuracy close to MUD-HoG (see Fig.~\ref{fig:acc_MNIST}) in the setups of 12.5\% and 20\% percentage of malicious clients.

\subsection{Confusion matrix}
\begin{figure}[h!]

    \centering
    \subfloat[MUD-HoG on MNIST]{
        \includegraphics[width=0.45\linewidth]{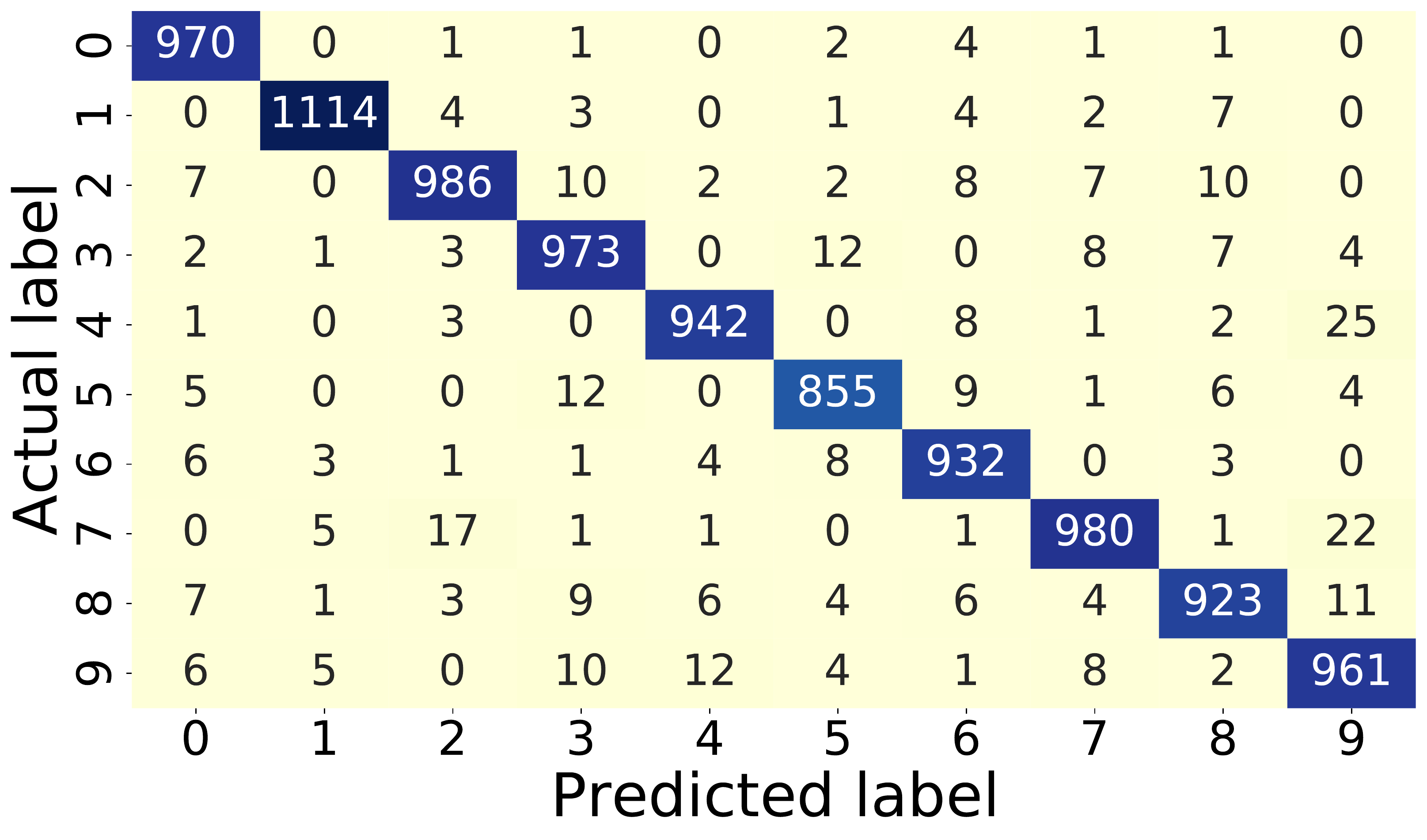}%
        \label{fig:confMatrixMudhogMNIST}
    }\quad
    \subfloat[FedAvg on MNIST]{
        \includegraphics[width=0.45\linewidth]{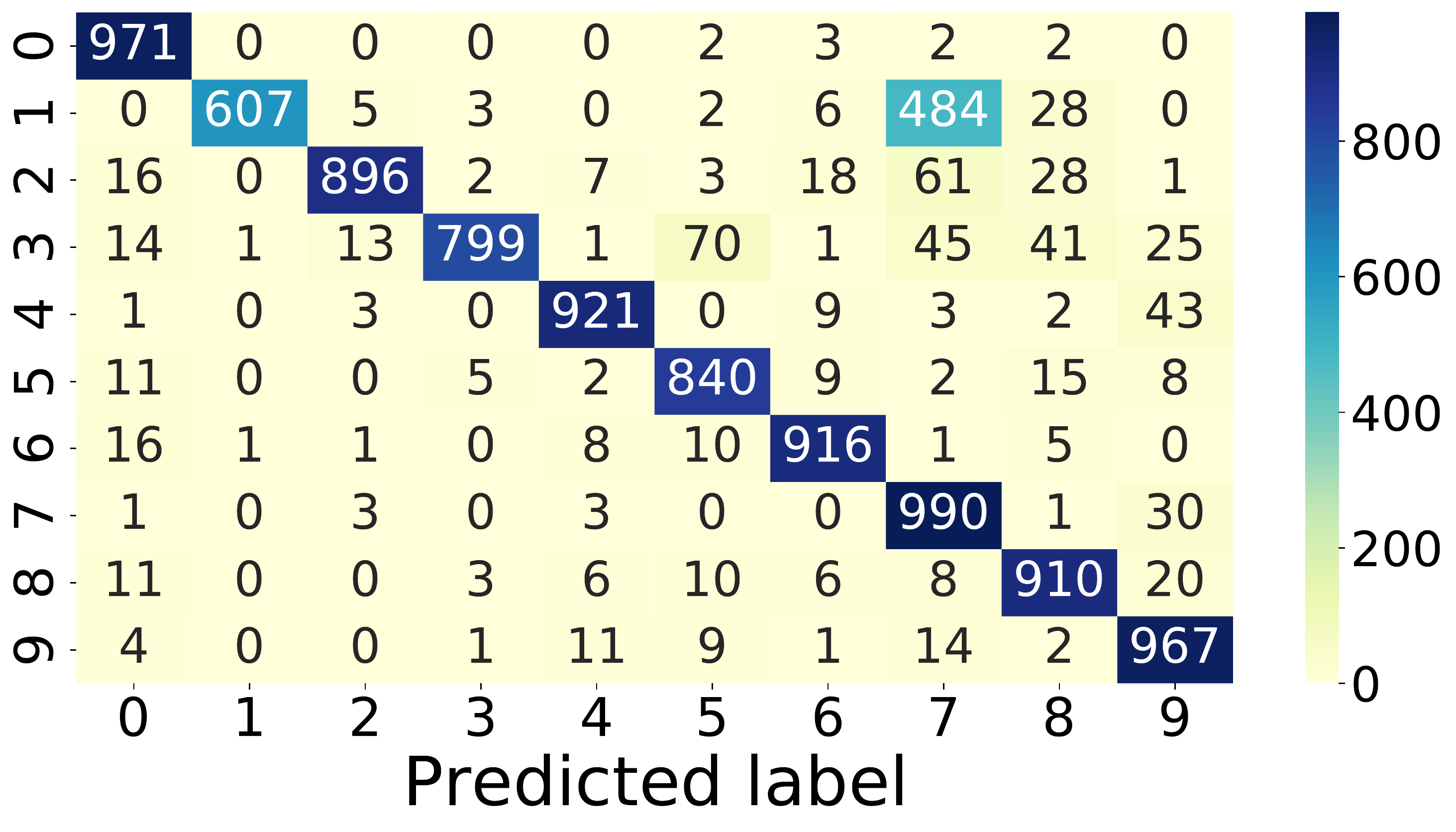}%
        \label{fig:confMatrixFedavgMNIST}
    }\\
    \bigskip
        \subfloat[MUD-HoG on Fashion-MNIST]{
        \includegraphics[width=0.45\linewidth]{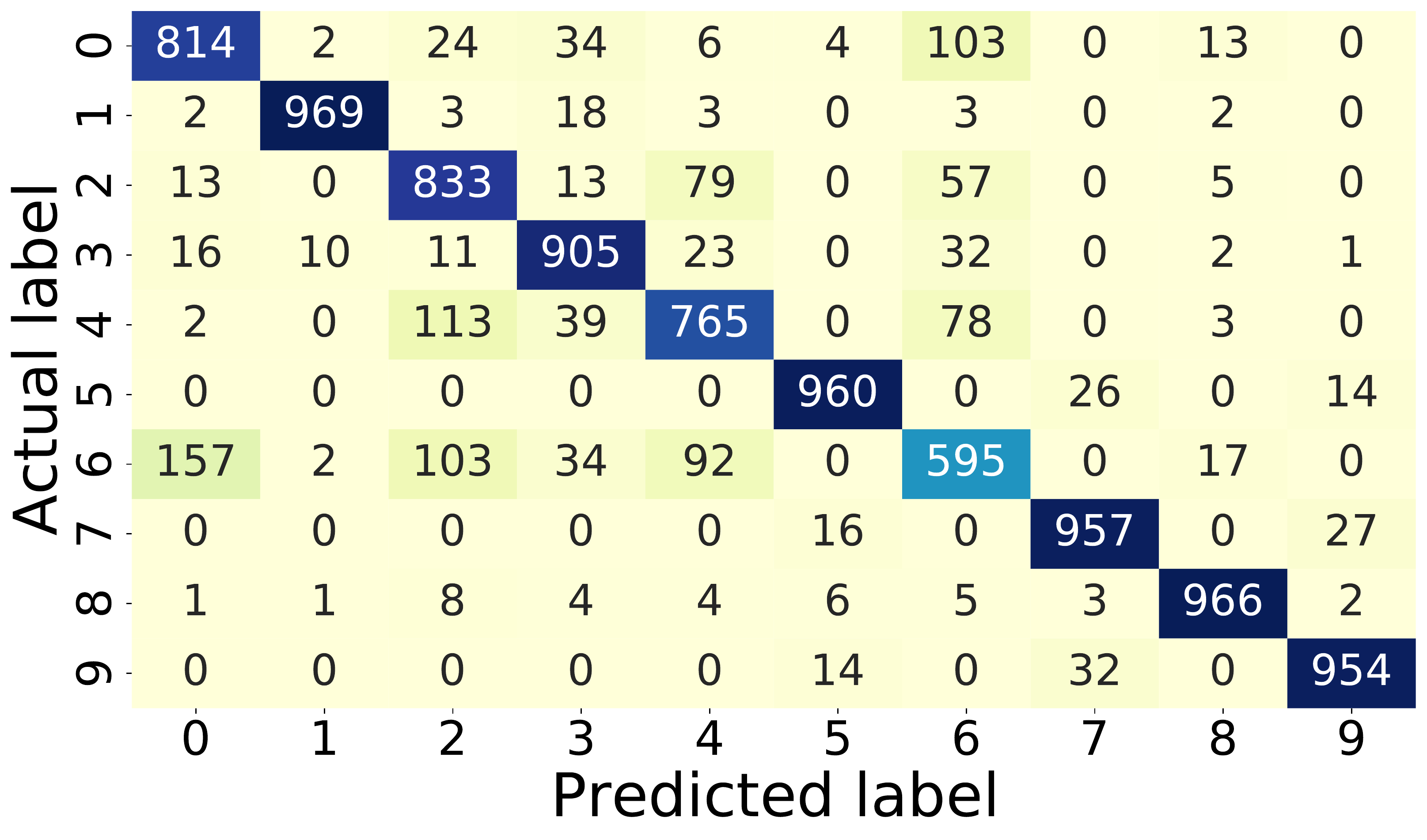}%
        \label{fig:confMatrixMudhogFashionMNIST}
    }
    \subfloat[FedAvg on Fashion-MNIST]{
        \includegraphics[width=0.45\linewidth]{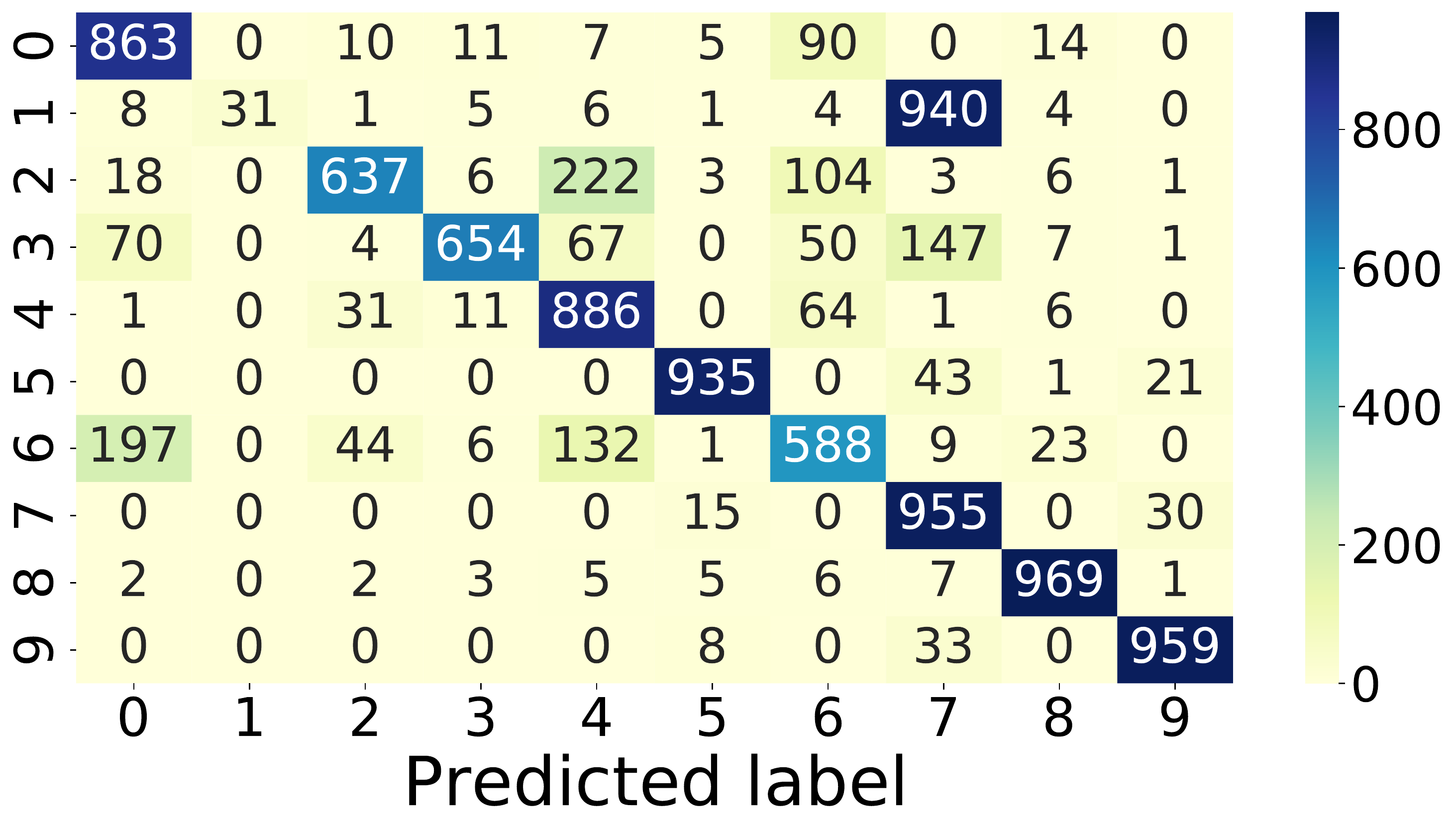}%
        \label{fig:confMatrixFedavgFashionMNIST}
    }
    
    \caption{Confusion matrices in {\it Series of Exp2} with 42.5\% malicious clients.}
     \label{fig:confMatrixMNIST}
    
\end{figure}

In Fig.~\ref{fig:confMatrixMNIST}, we show confusion matrices for MUD-HoG and FedAvg obtained from the completely trained model for MNIST and Fashion-MNIST datasets using a setup of series Exp2 with 42.5\% malicious clients. 
As multi-label-flipping attackers flip their local samples with source labels of "1", "2", and "3" to the target label "7", we can clearly see in parts (b) and (d) of Fig.~\ref{fig:confMatrixMNIST}, FedAvg confuses with several samples actually having the source labels as the target label while it is not the case for MUD-HoG. 
In addition, we see an interesting observation in part (d) of Fig.~\ref{fig:confMatrixMNIST}, where FedAvg completely fails as it predicts nearly all samples of source label "1" as the target label "7" (i.e., 940 samples of label "1" are predicted as label "7"). 

\newpage

\bibliographystyle{splncs04}
\bibliography{refer}

\begin{thebibliography}{10}
\providecommand{\url}[1]{\texttt{#1}}
\providecommand{\urlprefix}{URL }
\providecommand{\doi}[1]{https://doi.org/#1}

\bibitem{awan2021contra}
Awan, S., Luo, B., Li, F.: Contra: Defending against poisoning attacks in
  federated learning. In: European Symposium on Research in Computer Security
  (ESORICS). pp. 455--475. Springer (2021)

\bibitem{bagdasaryan2020backdoor}
Bagdasaryan, E., Veit, A., Hua, Y., Estrin, D., Shmatikov, V.: How to backdoor
  federated learning. In: International Conference on Artificial Intelligence
  and Statistics. pp. 2938--2948. PMLR (2020)

\bibitem{bhagoji2019analyzing}
Bhagoji, A.N., Chakraborty, S., Mittal, P., Calo, S.: Analyzing federated
  learning through an adversarial lens. In: International Conference on Machine
  Learning. pp. 634--643. PMLR (2019)

\bibitem{blanchard2017machine}
Blanchard, P., El~Mhamdi, E.M., Guerraoui, R., Stainer, J.: Machine learning
  with adversaries: Byzantine tolerant gradient descent. In: 31st International
  Conference on Neural Information Processing Systems. pp. 118--128 (2017)

\bibitem{cao2021fltrust}
Cao, X., Fang, M., Liu, J., Gong, N.Z.: Fltrust: Byzantine-robust federated
  learning via trust bootstrapping. In: ISOC Network and Distributed System
  Security Symposium (NDSS) (2021)

\bibitem{cao2021provably}
Cao, X., Jia, J., Gong, N.Z.: Provably secure federated learning against
  malicious clients. In: AAAI Conference on Artificial Intelligence. vol.~35,
  pp. 6885--6893 (2021)

\bibitem{chen2017distributed}
Chen, Y., Su, L., Xu, J.: Distributed statistical machine learning in
  adversarial settings: Byzantine gradient descent. ACM on Measurement and
  Analysis of Computing Systems  \textbf{1}(2),  1--25 (2017)

\bibitem{defazio2014saga}
Defazio, A., Bach, F., Lacoste-Julien, S.: Saga: A fast incremental gradient
  method with support for non-strongly convex composite objectives. In:
  Advances in neural information processing systems (2014)

\bibitem{fung2020limitations}
Fung, C., Yoon, C.J., Beschastnikh, I.: The limitations of federated learning
  in sybil settings. In: 23rd International Symposium on Research in Attacks,
  Intrusions and Defenses ($\{$RAID$\}$ 2020). pp. 301--316 (2020)

\bibitem{hard2018federated}
Hard, A., Rao, K., Mathews, R., Ramaswamy, S., Beaufays, F., Augenstein, S.,
  Eichner, H., Kiddon, C., Ramage, D.: Federated learning for mobile keyboard
  prediction. arXiv  (2018)

\bibitem{jiang2020federated}
Jiang, Y., Cong, R., Shu, C., Yang, A., Zhao, Z., Min, G.: Federated learning
  based mobile crowd sensing with unreliable user data. In: IEEE International
  Conference on High Performance Computing and Communications. pp. 320--327
  (2020)

\bibitem{khan2021federated}
Khan, L.U., Saad, W., Han, Z., Hossain, E., Hong, C.S.: Federated learning for
  internet of things: Recent advances, taxonomy, and open challenges. IEEE
  Communications Surveys \& Tutorials  \textbf{23}(3),  1759--1799 (2021)

\bibitem{lecun1998mnist}
LeCun, Y.: The mnist database of handwritten digits. http://yann. lecun.
  com/exdb/mnist/  (1998)

\bibitem{leroy2019federated}
Leroy, D., Coucke, A., Lavril, T., Gisselbrecht, T., Dureau, J.: Federated
  learning for keyword spotting. In: IEEE International Conference on
  Acoustics, Speech and Signal Processing. pp. 6341--6345 (2019)

\bibitem{li2019rsa}
Li, L., Xu, W., Chen, T., Giannakis, G.B., Ling, Q.: Rsa: Byzantine-robust
  stochastic aggregation methods for distributed learning from heterogeneous
  datasets. In: AAAI Conference on Artificial Intelligence. vol.~33, pp.
  1544--1551 (2019)

\bibitem{li2020learning}
Li, S., Cheng, Y., Wang, W., Liu, Y., Chen, T.: Learning to detect malicious
  clients for robust federated learning. arXiv  (2020)

\bibitem{liu2020fedvision}
Liu, Y., Huang, A., Luo, Y., Huang, H., Liu, Y., Chen, Y., Feng, L., Chen, T.,
  Yu, H., Yang, Q.: Fedvision: An online visual object detection platform
  powered by federated learning. In: AAAI Conference on Artificial
  Intelligence. vol.~34, pp. 13172--13179 (2020)

\bibitem{ma2021federated}
Ma, C., Li, J., Ding, M., Wei, K., Chen, W., Poor, H.V.: Federated learning
  with unreliable clients: Performance analysis and mechanism design. IEEE
  Internet of Things Journal  (2021)

\bibitem{mallah2021untargeted}
Mallah, R.A., Lopez, D., Farooq, B.: Untargeted poisoning attack detection in
  federated learning via behavior attestation. arXiv  (2021)

\bibitem{mao2021romoa}
Mao, Y., Yuan, X., Zhao, X., Zhong, S.: Romoa: Robust model aggregation for the
  resistance of federated learning to model poisoning attacks. In: European
  Symposium on Research in Computer Security (ESORICS). pp. 476--496. Springer
  (2021)

\bibitem{mcmahan2017communication}
McMahan, B., Moore, E., Ramage, D., Hampson, S., y~Arcas, B.A.:
  Communication-efficient learning of deep networks from decentralized data.
  In: Artificial intelligence and statistics. pp. 1273--1282. PMLR (2017)

\bibitem{nagalapatti2021game}
Nagalapatti, L., Narayanam, R.: Game of gradients: Mitigating irrelevant
  clients in federated learning. In: AAAI Conference on Artificial
  Intelligence. vol.~35, pp. 9046--9054 (2021)

\bibitem{nguyen2019new}
Nguyen, L.M., Nguyen, P.H., Richt{\'a}rik, P., Scheinberg, K., Tak{\'a}{\v{c}},
  M., van Dijk, M.: New convergence aspects of stochastic gradient algorithms.
  Journal of Machine Learning Research  \textbf{20},  1--49 (2019)

\bibitem{ozdayi2021defending}
Ozdayi, M.S., Kantarcioglu, M., Gel, Y.R.: Defending against backdoors in
  federated learning with robust learning rate. In: AAAI Conference on
  Artificial Intelligence. vol.~35, pp. 9268--9276 (2021)

\bibitem{schubert2017dbscan}
Schubert, E., Sander, J., Ester, M., Kriegel, H.P., Xu, X.: Dbscan revisited,
  revisited: why and how you should (still) use dbscan. ACM Transactions on
  Database Systems (TODS)  \textbf{42}(3) (2017)

\bibitem{sun2019can}
Sun, Z., Kairouz, P., Suresh, A.T., McMahan, H.B.: Can you really backdoor
  federated learning? arXiv  (2019)

\bibitem{tolpegin2020data}
Tolpegin, V., Truex, S., Gursoy, M.E., Liu, L.: Data poisoning attacks against
  federated learning systems. In: European Symposium on Research in Computer
  Security (ESORICS). pp. 480--501. Springer (2020)

\bibitem{Wan2021RobustFL}
Wan, C.P., Chen, Q.: Robust federated learning with attack-adaptive
  aggregation. ArXiv  \textbf{abs/2102.05257} (2021)

\bibitem{wang2020attack}
Wang, H., Sreenivasan, K., Rajput, S., Vishwakarma, H., Agarwal, S., Sohn,
  J.y., Lee, K., Papailiopoulos, D.: Attack of the tails: Yes, you really can
  backdoor federated learning. arXiv  (2020)

\bibitem{wu2020federated}
Wu, Z., Ling, Q., Chen, T., Giannakis, G.B.: Federated variance-reduced
  stochastic gradient descent with robustness to byzantine attacks. IEEE
  Transactions on Signal Processing  \textbf{68},  4583--4596 (2020)

\bibitem{fashionMNIST}
Xiao, H., Rasul, K., Vollgraf, R.: Fashion-mnist: a novel image dataset for
  benchmarking machine learning algorithms (2017)

\bibitem{xie2021crfl}
Xie, C., Chen, M., Chen, P.Y., Li, B.: Crfl: Certifiably robust federated
  learning against backdoor attacks. In: International Conference on Machine
  Learning. pp. 11372--11382. PMLR (2021)

\bibitem{xie2018generalized}
Xie, C., Koyejo, O., Gupta, I.: Generalized byzantine-tolerant sgd. arXiv
  (2018)

\bibitem{xie2019zeno}
Xie, C., Koyejo, S., Gupta, I.: Zeno: Distributed stochastic gradient descent
  with suspicion-based fault-tolerance. In: International Conference on Machine
  Learning. pp. 6893--6901. PMLR (2019)

\bibitem{yin2018byzantine}
Yin, D., Chen, Y., Kannan, R., Bartlett, P.: Byzantine-robust distributed
  learning: Towards optimal statistical rates. In: International Conference on
  Machine Learning. pp. 5650--5659. PMLR (2018)

\end{thebibliography}
\end{document}